# Infrared, terahertz and microwave spectroscopy of the soft and central modes in $Pb(Mg_{1/3}Nb_{2/3})O_3$

D. Nuzhnyy, J. Petzelt, V. Bovtun, M. Kempa, S. Kamba, J. Hlinka,
Institute of Physics, Czech Academy of Sciences
Na Slovance 2, 18221 Prague 8, Czech Republic
B. Hehlen,
Laboratoire Charles Coulomb (L2C), UMR 2251,
CNRS-Université de Montpellier, F-34095 Montpellier, France

## Abstract

From the new infrared (IR) reflectivity and time-domain terahertz (THz) spectra combined with available high-frequency dielectric data above the MHz range in a broad temperature range of 10-900 K, a full picture of the soft and central mode behavior in the classical relaxor ferroelectric $Pb(Mg_{1/3}Nb_{2/3})O_3$ (PMN) is suggested. A detailed comparison is given with the recent hyper-Raman spectroscopy data (Hehlen et al. *Phys. Rev. Lett.* **117,** 155501 (2016)), and also with other available experiments based on inelastic light and neutron scattering. It is revealed that each type of experiment provides slightly different data. The closest agreement is with the hyper-Raman data, both techniques yield the same number of soft-mode components and the same high-temperature softening towards the temperature $T^* \approx 400$ K. In addition to evaluation of the IR-THz data using fitting with standard factorized form of the dielectric function, we performed a successful fitting of the same data using the effective medium approach (EMA), originally based on the assumption that the mesoscopic structure of PMN consists of randomly oriented uniaxially anisotropic polar nanodomains (PNDs) with somewhat harder TO polar modes in the direction along the local PND dipole (*Phys. Rev. Lett.* **96,** 027601 (2006)). Evaluation using the Bruggeman EMA modelling has been successfully applied in the entire investigated temperature range. These results suggest that the response perpendicular to the local dipole moment, at high temperatures induced by random fields rather than PNDs, undergoes a classical softening from high temperatures with permittivity obeying the Curie-Weiss law, $\varepsilon_\perp = C/(T-T_C)$, $C = 1.7 \times 10^5$ K and $T_C = 380$ K, whereas the response parallel to it shows no softening. Below the Burns temperature ~620 K, a GHz relaxation ascribed to flipping of the PNDs emerges from the soft mode response, slows down and broadens, remaining quite strong towards the cryogenic temperatures, where it can be assigned to fluctuations of the PND boundaries.

## I. INTRODUCTION

In ideal displacive proper ferroelectrics, IR active soft phonon mode (SM) induces the ferroelectric phase transition and its temperature-dependent dielectric strength fully accounts for the Curie-Weiss dielectric anomaly near the phase transition temperature $T_C$. Most of displacive ferroelectrics, however, exhibit only partial phonon softening towards $T_C$ and close to $T_C$ an additional soft overdamped excitation appears below the SM response, called central mode (CM), which can have comparable or even higher dielectric strength (contribution to static permittivity) than the SM [1]. The SM (and CM) behavior of relaxor ferroelectrics and particularly of the classical perovskite relaxor $Pb(Mg_{1/3}Nb_{2/3})O_3$ (PMN) represents a non-trivial task for lattice dynamics studies already for about 20 years, as it has become clear that, unlike in displacive ferroelectrics, their dielectric strengths cannot explain the large dielectric anomaly, dispersive down to very low frequencies [1]. Moreover, the significance and physical relevance of the SM behavior in relaxors is not so evident, since it is not connected with any macroscopic phase transition. Therefore the softening of the SM is not expected to be complete (it may be connected with some local phase transition) and the concept of CM is not quite well defined, because the lower-frequency dielectric relaxations, which are generally ascribed to appearance of dynamic polar nanoregions or nanodomains, are usually treated separately from the CM and SM response. Suitable techniques for the SM and CM studies are inelastic neutron scattering (INS), inelastic light scattering as Raman, Brillouin and hyper-Raman spectroscopy (HRS), and Fourier-transform infrared (IR), time-domain terahertz (THz) and microwave spectroscopy, which offer the high-frequency dielectric response above ~MHz, in combination with the standard low-frequency capacitance spectroscopy, many times published as most recently reviewed by Bokov and Ye [2]. The rich INS studies before 2011 have been thoroughly reviewed by Cowley et al. [3] and here we will compare only the recent and for SM and CM spectroscopy relevant INS results [3-9], Raman [10-13], Brillouin [14] and HRS [15-17] studies with the dielectric response from the $10^6$ to$10^{13}$ Hz presented and analyzed here. Particularly, a detailed comparison of the HRS results with those of IR-THz spectroscopy is of interest, because the selection rules for HRS and IR are similar [15] – the SM should be active (unlike in Raman) even for the average simple cubic perovskite structure of PMN, space group $Pm\overline{3}m$ with one formula unit per unit cell ($Z$ = 1).

The main difficulty with the SM studies in PMN is that, on cooling from high temperatures, the low-frequency SM becomes overdamped in the most interesting temperature range between the Burns temperature $T_B \approx 620$ K and the temperature $T^* \approx 400$ K [17]. Typically, it is assumed that either the dynamic polar nanoregions, embedded into the paraelectric matrix, or the dynamic polar nanodomains (PNDs), separated by more or less sharp domain boundaries from the neighbouring PND's, are formed at $T_B$ while below the temperature $T^*$ they begin to freeze. The difference between the polar nanoregions within a paraelectric matrix and the PNDs, occupying the whole volume, was discussed in Ref. [18]. Even if the preferred picture is not yet commonly accepted, here we will prefer the picture of PNDs, which better corresponds to our results. Namely, in this case below $T_B$, the complete dielectric response consists only of the contributions of PNDs (their bulk and boundaries) and the response of the paraelectric matrix can be neglected. Below the complete freezing temperature $T_f \approx 200$-$220$ K, where the relatively stable ferroelectric phase (still consisting of static PNDs) can be induced by an external electric field, the SM response becomes underdamped, but splits into at least two components [3,17]. In inelastic neutron and light scattering experiments, the overdamped SM appears as a central peak (i.e. CM), whose half-width represents the characteristic frequency (corresponding to the frequency of the dielectric loss peak in the dielectric response), which is, however, hard to determine accurately, since it is usually overlapped with the elastic central peak (Bragg, Rayleigh). This problem is avoided in the dielectric response, where the loss peak of the CM has always a finite frequency, but the typical frequency range of this peak ($10^{11}$ – $10^{12}$ Hz or 3 – 30 cm$^{-1}$), appearing frequently close but below the standard THz range, is rather hardly accessible.

The first temperature-dependent SM studies of PMN using IR and THz spectroscopy were performed by Bovtun et al. [19], who joined together these data with the lower-frequency dielectric data and assigned them in accordance with the general scheme of the dielectric spectra in relaxors published somewhat earlier [20]. In this scheme the CM appears below $T_B$ where it starts to separate from the SM response in the $10^{11}$ Hz range, slows down as a relaxation to very low (~mHz) frequencies at the temperature of low-frequency permittivity maximum (~240 K), and is assigned to the dynamics of PNDs. The data on SM and CM in PMN were already reviewed and compared with other relaxors and ferroelectrics in Ref. [1]. However, here we will use the concept of CM in PMN rather in a more restricted sense, as the overdamped component of the SM response, if the response consists of more than one component. The GHz and lower-frequency dispersion caused by PND dynamics will be called relaxational dispersion. At the same time, the available IR-THz, Raman and INS data were also compared with the first-principles lattice dynamics calculations of variously ordered PMN supercells using the frozen phonon method by Prosandeev et al. [21]. Due to the high absorption in the THz range, the time-domain THz transmission data on PMN crystal plate samples were obtained only below ~160 K. To broaden the temperature range up to ~900 K, Kamba et al. [22] studied the IR transmission of a PMN thin film (prepared by chemical solution deposition on a sapphire substrate, thickness ~500 nm), from which they could evaluate the SM and CM response, which at low temperatures agreed with the single crystal response. The data on the whole available dielectric response of the PMN single crystals, ceramics and thin films were summarized by Bovtun et al. [23], concluding that concerning the SM response there was no substantial difference among the types of the sample.

The IR-THz reflectivity spectra of PMN revealed more complex features than expected from the simple cubic structure, for which only three $F_{1u}$ IR active transverse optical (TO) phonon modes are expected (see [24] for a detailed review and analysis of the IR spectra of ferroelectric and relaxor ferroelectric perovskites). Therefore Hlinka et al. [25] suggested that the PNDs show locally an anisotropic dielectric response, differing for the electric field $\boldsymbol{E}$ parallel and perpendicular to the local polarization $\boldsymbol{P}_l$ of PNDs, which splits the effective phonon response. To model such spectra, an effective medium approximation (EMA) was used with a Bruggeman model, mixing statistically the $\boldsymbol{E}||\boldsymbol{P}_l$ (1/3 volume fraction) and $\boldsymbol{E}\perp\boldsymbol{P}_l$ response (2/3 volume fraction) and a very good fit was obtained for the 300 K IR reflectivity spectra. This approach could explain the two-mode behavior, i.e. splitting of all the three $F_{1u}$ phonons below $T_B$ into the $E$ and $A_1$ symmetry modes.

However, recent INS as well as inelastic light scattering and IR data have revealed two modes in the SM range even above $T_B$. By comparison of the HRS with the Raman data [13] it was suggested that the second component of the SM near 45 $cm^{-1}$ in the HRS data [17] is the $F_{2g}$ mode activated in the Raman spectra due to the local doubling of the unit cell caused by the local one-to-one B-site ordering in the $ABO_3$ perovskite structure (small chemical clusters of the $Fm\bar{3}m$, $Z$ = 2 structure). This mode was weakly seen in the low-temperature IR spectra, in addition to the SM doublet [22,23], and it was clearly detected in the whole temperature range 30 – 900 K also in the HRS experiment [17]. Since the selection rules concerning SM for the HRS and IR are expected to be the same [15] and the corresponding spectra look qualitatively similar, it is of interest to compare both responses in the whole temperature range more quantitatively. For that in this work we have complemented our previous IR data by new THz measurements with a thin plate of dense PMN ceramics (described in [26]) and carried out the IR reflectivity measurements on single crystals, both up to 900 K. These data were analyzed together with our earlier data in the lower-frequency range [19,20,23] and compared with the HRS data [17].

## II. EXPERIMENT AND EVALUATION

To cover the 0.2 – 21 THz frequency range (~6 - 700 $cm^{-1}$ wavenumber range) we have used two techniques: time-domain THz transmission (0.2 – 1 THz) and IR reflectivity (1 – 21

THz) spectroscopy. The time-domain THz transmission spectrometer is based on a Ti:sapphire femtosecond laser source [27]. The THz pulses were generated by an interdigitated photoconducting GaAs switch. Electro-optic sampling with a plate of (110) ZnTe crystal was used to detect the transmitted THz pulses. IR reflectivity spectra were obtained using the Fourier-transform IR spectrometer Bruker IFS 113v with deuterated triglycine sulfate pyroelectric detector. We used an Optistat continuous He-flow cryostat with mylar windows for the low-temperature THz measurements and commercial high-temperature cell SPECAC P/N 5850 for the high-temperature range in both experiments. THz measurements were performed on the dense polished plane-parallel ceramic plate (with a small porosity of 4.3%) [26], thickness 43 μm, and the high-temperature IR reflectivity spectra were obtained on the polished thick PMN crystal used in Ref. [21]. Low-temperature IR reflectivity data were taken from the previous experiment on the PMN crystal [19,21]. The experimental data performed here in the 0.2 – 30 THz frequency range were compared with the data on ceramics and crystals from our earlier measurements [19,21,22]. We found that all the data are well compatible. From the combined earlier and new experimental data we have chosen data for simultaneous fitting the IR reflectivity and THz complex dielectric spectra in a broad temperature range 10–900 K. The fitting was performed using the standard factorized formula of the generalized multi-oscillator dielectric function [1]:

$$\varepsilon(\omega) = \varepsilon_\infty \prod_j \frac{\omega_{LOj}^2 - \omega^2 + i\omega\gamma_{LOj}}{\omega_{TOj}^2 - \omega^2 + i\omega\gamma_{TOj}} \quad (1)$$

related to the normal reflectivity by

$$R(\omega) = \left| \frac{\sqrt{\varepsilon(\omega)} - 1}{\sqrt{\varepsilon(\omega)} + 1} \right|^2 \quad (2)$$

where $\varepsilon_\infty$ is the high-frequency permittivity resulting from the electronic absorption processes much above the polar phonons, $\omega$ is the linear frequency or wavenumber, $\omega_{TOj}$ and $\gamma_{TOj}$ is the transverse optic (TO) frequency and damping parameter of the $j$-th oscillator, respectively, and $\omega_{LOj}$ and $\gamma_{LOj}$ is its respective longitudinal optic (LO) frequency and damping parameter. In formula (1) each generalized oscillator is characterized by 4 parameters compared with the classical damped oscillator formula with 3 parameters.

Experimental details concerning the complex dielectric spectra of PMN crystals between 3 mHz and 75 GHz were published earlier [19,20,28-30] and the data were joined together in Ref. [23]. Here we have joined them with our THz data and fitted the data above ~1 MHz independently of the previous IR fits with a sum of Cole-Cole relaxations plus one damped harmonic THz oscillator in the 100–500 K temperature range, where the data above 1 MHz are available:

$$\varepsilon(\omega) = \varepsilon_{ph} + \sum_j \frac{\Delta\varepsilon_j}{1 + \left( {i\omega}/{\omega_{Rj}} \right)^{1-\alpha_j}} + \frac{\Delta\varepsilon_{HO}\omega_{HO}^2}{\omega_{HO}^2 - \omega^2 + i\omega\gamma_{HO}} \quad (3)$$

where $\varepsilon_{ph}$ is the high-frequency contribution to the dielectric function obtained from the fits to IR reflectivity above ~50 cm$^{-1}$, $\Delta\varepsilon_j$ is the dielectric strength of the $j$-th Cole-Cole relaxation, $\omega_{Rj}$ is the mean relaxation frequency of the $j$-th relaxation and $0 \leq \alpha_j \leq 1$ describes the degree of broadening of the $j$-th Cole-Cole relaxation in the loss spectra ($\alpha_j = 0$ corresponds to the Debye relaxation, $\alpha_j \rightarrow 1$ describes the frequency-independent loss spectra, i.e. infinitely broad uniform distribution of Debye relaxations), $\Delta\varepsilon_{HO}$, $\omega_{HO}$ and $\gamma_{HO}$ is the dielectric strength, linear frequency and damping of the THz oscillator, respectively.

It is not possible to fit the whole spectra together including the IR range with a single model formula, because the relaxations yield unphysically high losses in the high-frequency range (far above the loss peaks) and do not fulfil the sum rule, see e.g. [31]. Therefore the fits with relaxations can be used at most up to the THz frequency range.

## III. IR-THz DATA - RESULTS

In Fig. 1 we plot the IR reflectivity spectra of PMN in the whole measured temperature range 10–900 K together with the calculated low-frequency part from the THz transmission data and fits using Eqs. 1 and 2 with up to 11 TO and LO oscillators. Below 300 K we have the measured data only below 650 cm$^{-1}$, limited by the transparency spectral range of the polyethylene windows of the cryostat, and the $\varepsilon_\infty$ parameter was taken as temperature independent (model (i)).

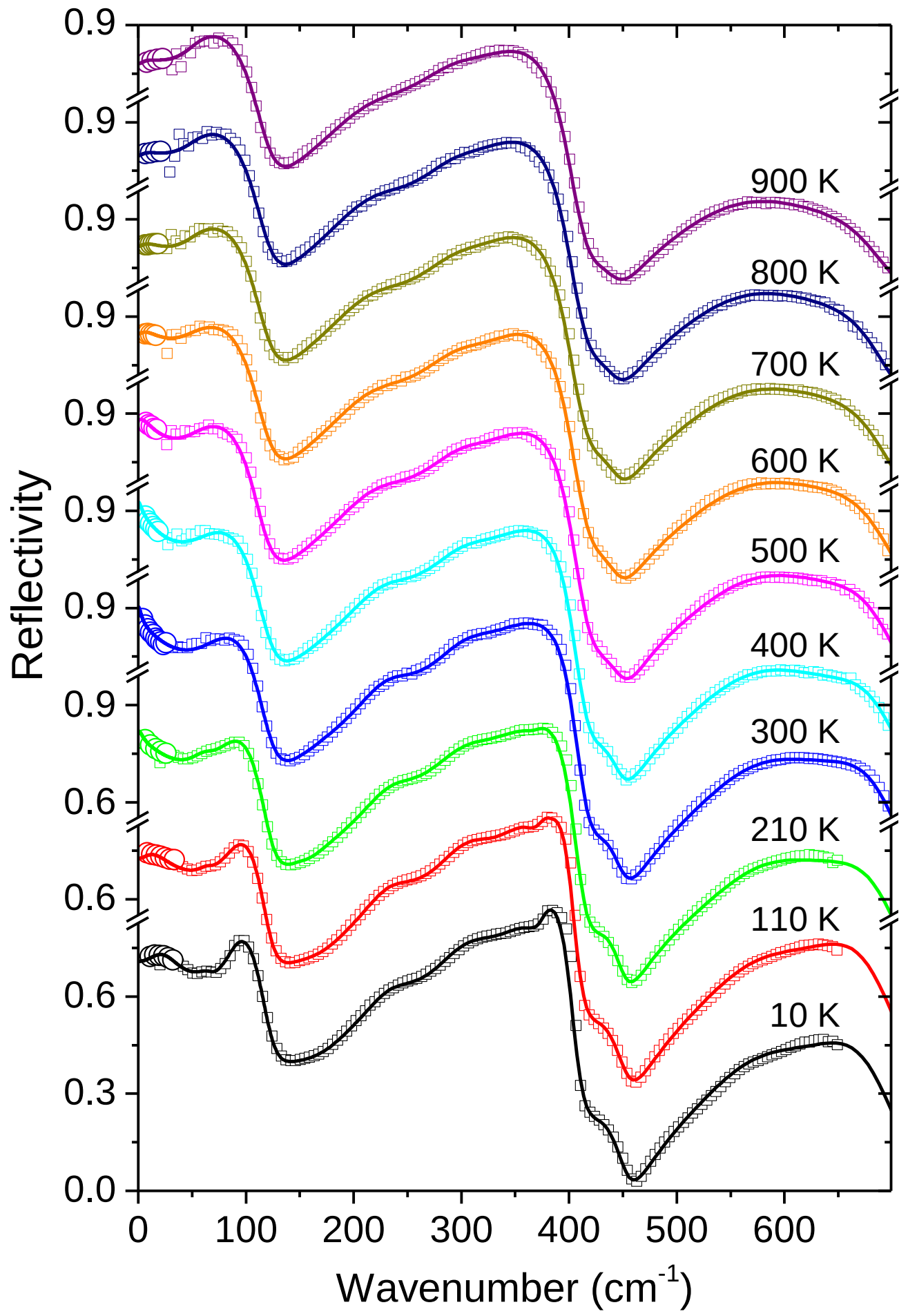

FIG. 1. Measured unpolarized IR reflectivities (squares) of the PMN crystal together with the calculated THz data (circles) and the factorized formula fits using the Eqs. 1 and 2 (full lines) in the 10–900 K range (model (i)). The fitted reflectivities agree with the data within the limits of experimental accuracy.

The real permittivity spectra calculated from the fits (together with the THz data) are plotted in the log-wavenumber scale in Fig. 2.

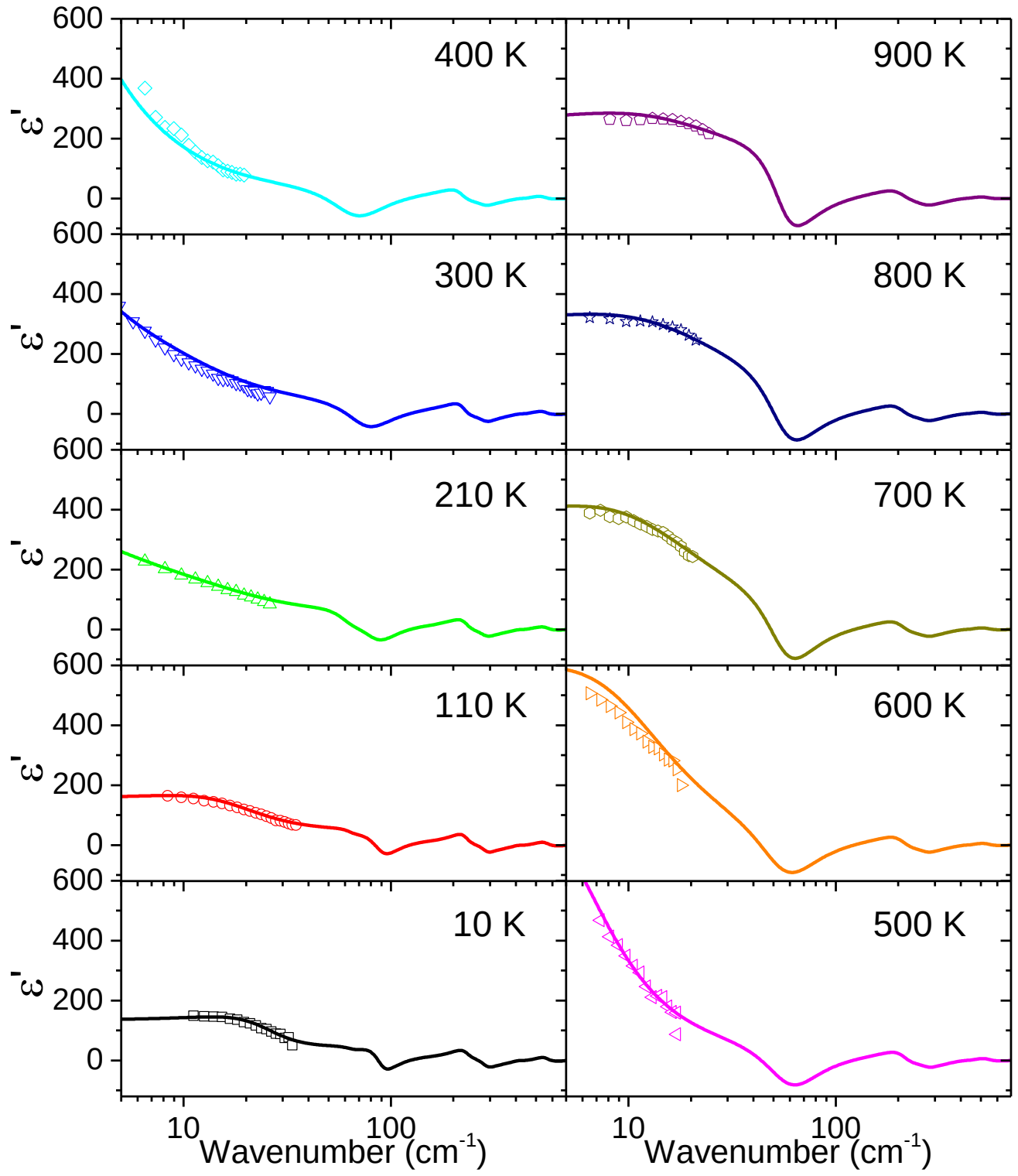

FIG. 2. Permittivity spectra (full lines) calculated from the reflectivity fits shown in Fig. 1 (model (i)) and the THz data (symbols). Note the log-wavenumber scale.

Instead of the usually presented dielectric loss spectra $\varepsilon''(\omega)$, in Fig. 3 we prefer to show the real optical conductivity data $\sigma'(\omega)$ in the linear wavenumber scale, related by $\sigma'(\omega) = 2\pi\varepsilon_0\omega\varepsilon''(\omega)$. The conductivity spectra have the advantage of fulfilling the well-known oscillator-strength sum rule, which relates the integral over the conductivity spectra to the total electric volume-unit charge which takes part in all the dynamic motion of ions and electrons in the material under question. In the case of one type of charges *e* with concentration *N* and mass *m* it sounds [32]

$$\int_0^\infty \sigma'(\omega)d\omega = \frac{\pi}{2}\frac{Ne^2}{m} \tag{4}$$

It is expected to be essentially independent of temperature even if the system undergoes some phase transitions. Another advantage of the conductivity spectra is that, for a classical damped oscillator response, the conductivity maximum corresponds to the oscillator frequency $\omega_{HO}$ for any damping, including overdamped oscillators ($\gamma_{HO} > 2\omega_{HO}$) [16]. It differs from the maximum of the oscillator response in the loss spectra $\varepsilon''(\omega)$, which for low damping corresponds approximately to the same frequency, but for high damping its frequency decreases below $\omega_{HO}$, and for overdamped oscillators it approaches the frequency $\omega_{HO}^2/\gamma_{HO}$.

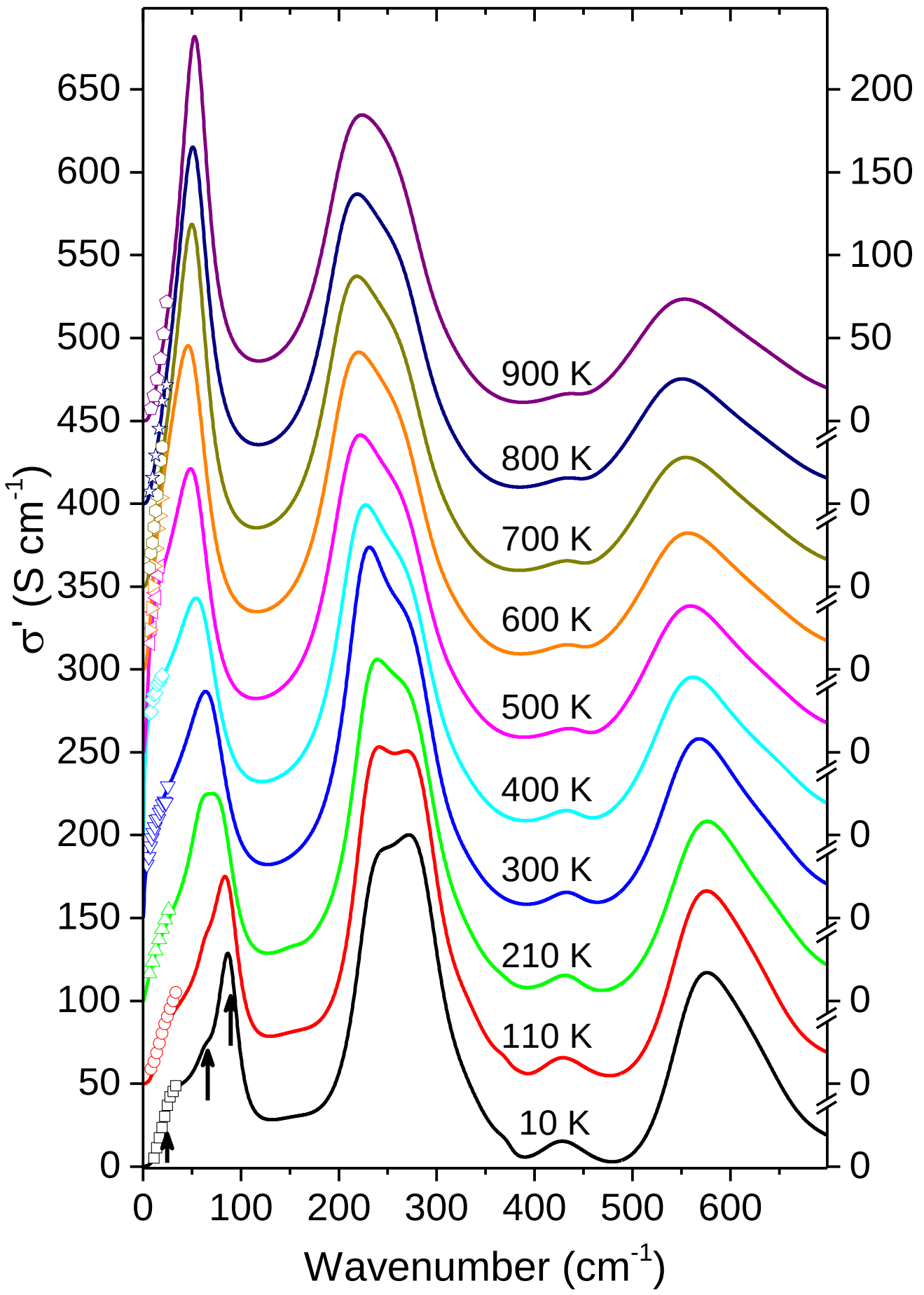

FIG. 3. IR conductivity spectra (full lines) calculated from the fits in Fig. 1 (model (i)), including the THz data (symbols). Each spectrum is shifted by 50 S cm$^{-1}$ with respect to the previous one and the values on the left-hand *y* axis match the spectrum at 10 K, while on the right-hand y-axis the proper scale is adapted for all the spectra. Arrows mark the three SM features analyzed below (see Fig. 6).

## IV. COMPARISON WITH THE HYPER-RAMAN SCATTERING

The recent HRS data [17] present the most accurate spectroscopic data below ~100 cm$^{-1}$ in the broad temperature range 30 – 900 K and suggest a new assignment of the observed SM doublet and triplet above and below ~300 K, respectively. The authors [17] suggest that the mode lying near 45 cm$^{-1}$, which is present at all temperatures and is only weakly temperature dependent, is the $F_{2g}$ Raman mode activated due to a local doubling of the unit cell (local B-site one-to-one ordering) [10,11,13]. According to this picture, above ~300 K the SM seen below 40 cm$^{-1}$ is only a singlet which appears to be bilinearly coupled with the $F_{2g}$ mode due to the B-site disorder, which relaxes the selection rules. Below ~300 K the SM splits into two components due to the uniaxial anisotropy of the PNDs (almost frozen), as already earlier suggested from the analysis of the IR reflectivity based on EMA [25], with the $F_{2g}$ mode remaining between the SM components as the third phonon mode below 100 cm$^{-1}$.

Our IR-THz data reveal qualitatively the same picture. For fitting the range below 100 cm$^{-1}$ using Eqs. 1 and 2 (model (i)) we have quite independently used two TO modes at $T \geq 300$ K and three TO modes for $T < 300$ K. For a more quantitative comparison we have used another fitting model of the IR data, which is directly comparable to the HRS data: the low-frequency part below 100 cm$^{-1}$ was fitted with two (for $T \geq 400$ K) or three (for $T \leq 300$ K) additive classical damped harmonic oscillators and the remaining higher-frequency part of the spectra was additively fitted with the factorized form of generalized oscillators, Eq. 1, with the fitting parameters generally differing from those used in model (i):

$$\varepsilon(\omega) = \sum_{k} \frac{\Delta\varepsilon_k \omega_k^2}{\omega_k^2 - \omega^2 + i\omega\gamma_k} + \varepsilon_\infty \prod_{j} \frac{\omega_{LOj}^2 - \omega^2 + i\omega\gamma_{LOj}}{\omega_{TOj}^2 - \omega^2 + i\omega\gamma_{TOj}} \quad (5)$$

where the first right-hand sum was used for fitting the range below 100 cm$^{-1}$ and all the parameters were introduced above in Eqs. 1 and 3. First we have fixed the same frequencies and dampings as used for fitting the HRS spectra and fitted only the oscillator strengths, which have no relation to the HRS mode strengths. However, the fits were not of good quality and are not shown. In the next step we have fitted all the parameters of the additive low-frequency oscillators (model (ii)) and in Figs. 4 and 5 we plot the low-frequency part below 150 cm$^{-1}$ of the fitted reflectivity and of the complex dielectric response with the THz data and our fits from Figs. 1-3.

In Fig. 4 we show the spectra for $T \leq 300$ K and in Fig. 5 for $T \geq 400$ K, but one can see that the fits are still not quite perfect. In Fig. 6 we have plotted our low-frequency fitting parameters of the additive oscillators (frequencies and dampings, model (ii)) for all the temperatures and compared them with those from the HRS spectra fits [17].

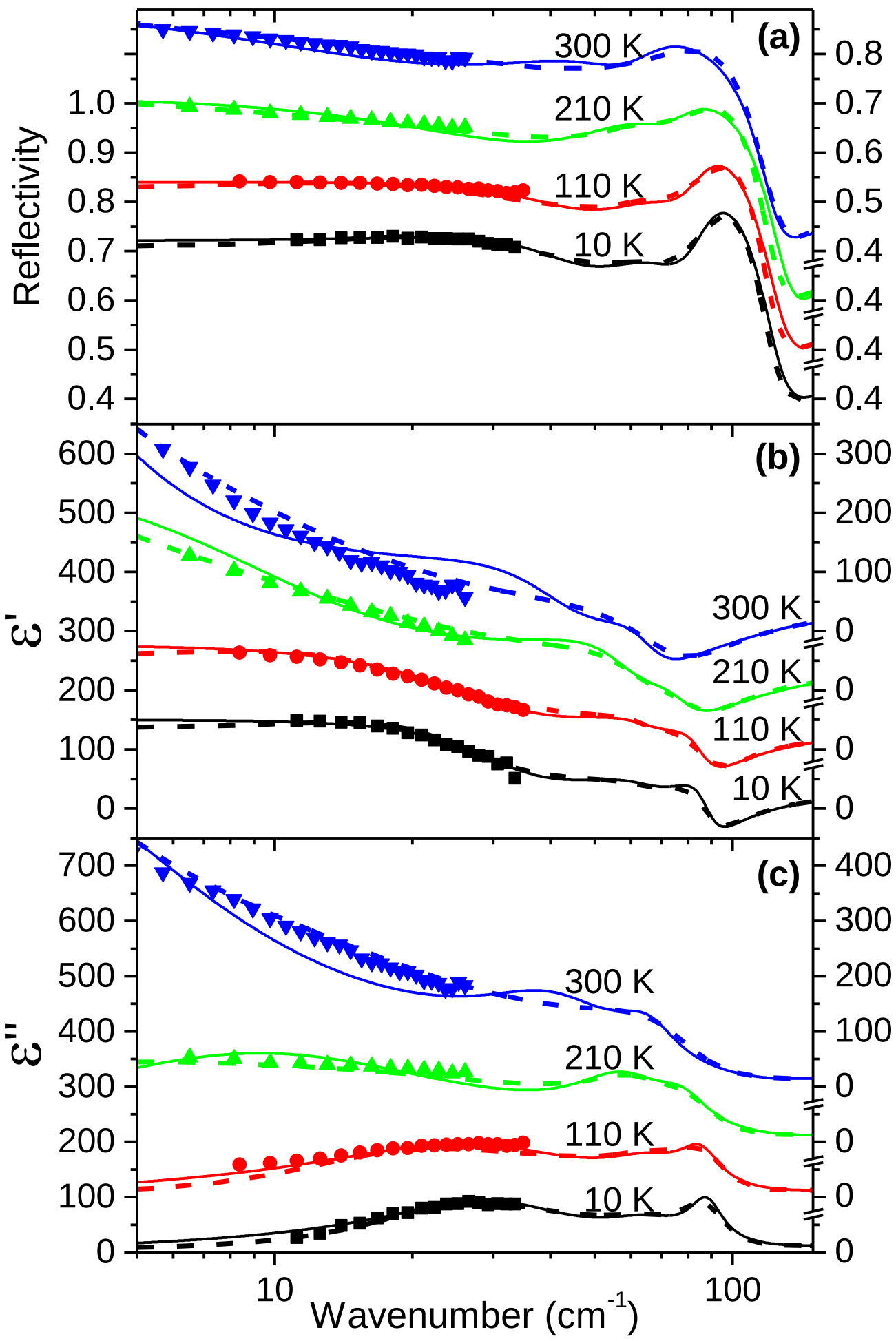

FIG. 4. Low-frequency and low-temperature part of the IR fits using the three additive oscillator fits in the log-wavenumber scale (model (ii), see the text) – full lines, as compared with the THz data – symbols and the fits from Figs. 1-3 (model (i)) – dashed lines. The values on the left-hand y-axis match the spectrum at 10 K, while on the right y-axis the proper scale is adapted for all the spectra.

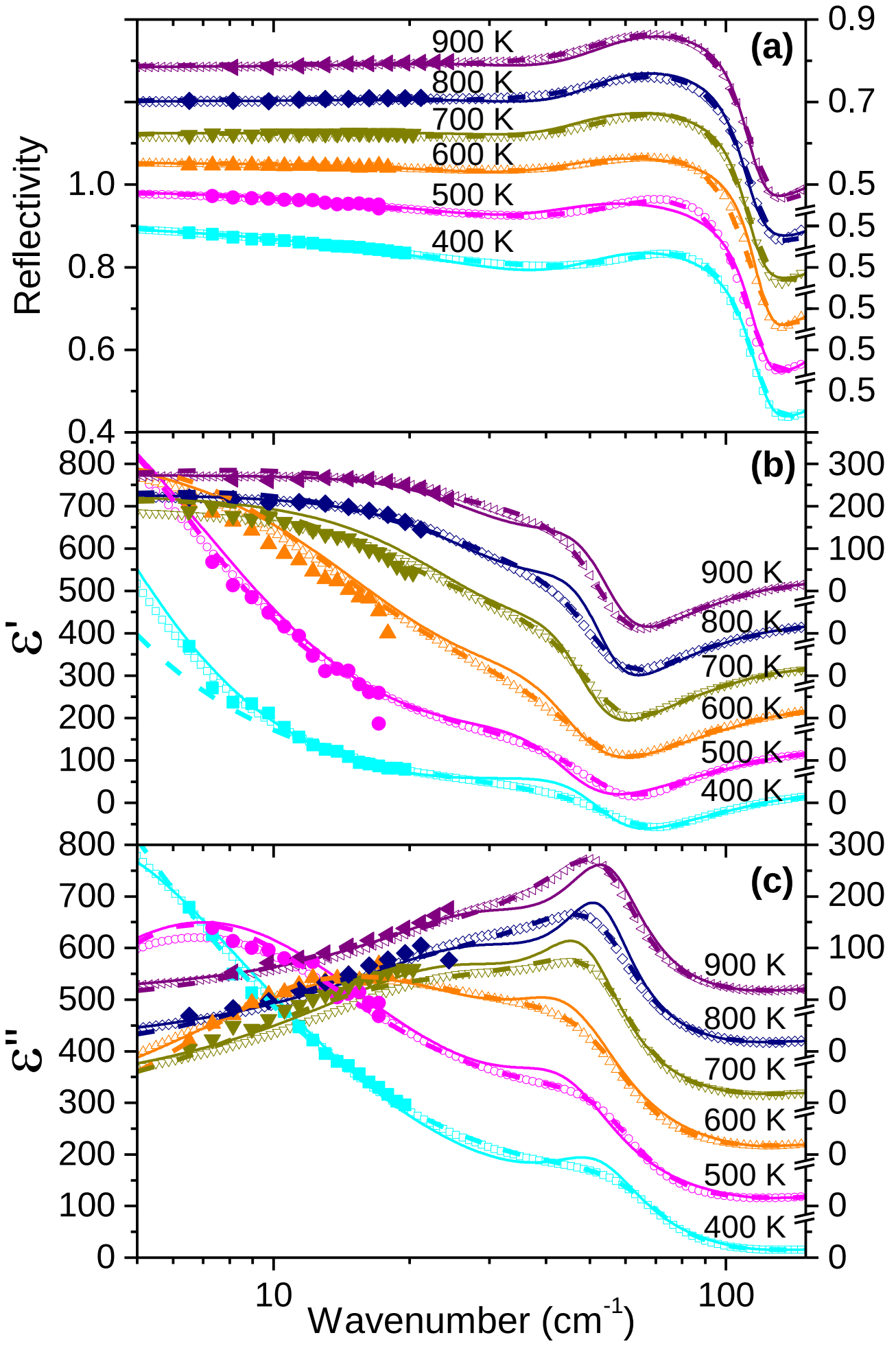

FIG. 5. Low-frequency and high-temperature part of the IR fits using the two additive oscillator fits in the log-wavenumber scale (model (ii)) – full lines, as compared with the THz data – full symbols, fits from Fig. 1 – dashed lines, and fits using imaginary coupling between the oscillators (model (iii) - see text) – open symbols. The values on the left-hand y-axis match the spectrum at 400 K, while on the right-hand y-axis the proper scale is adapted for all the spectra.

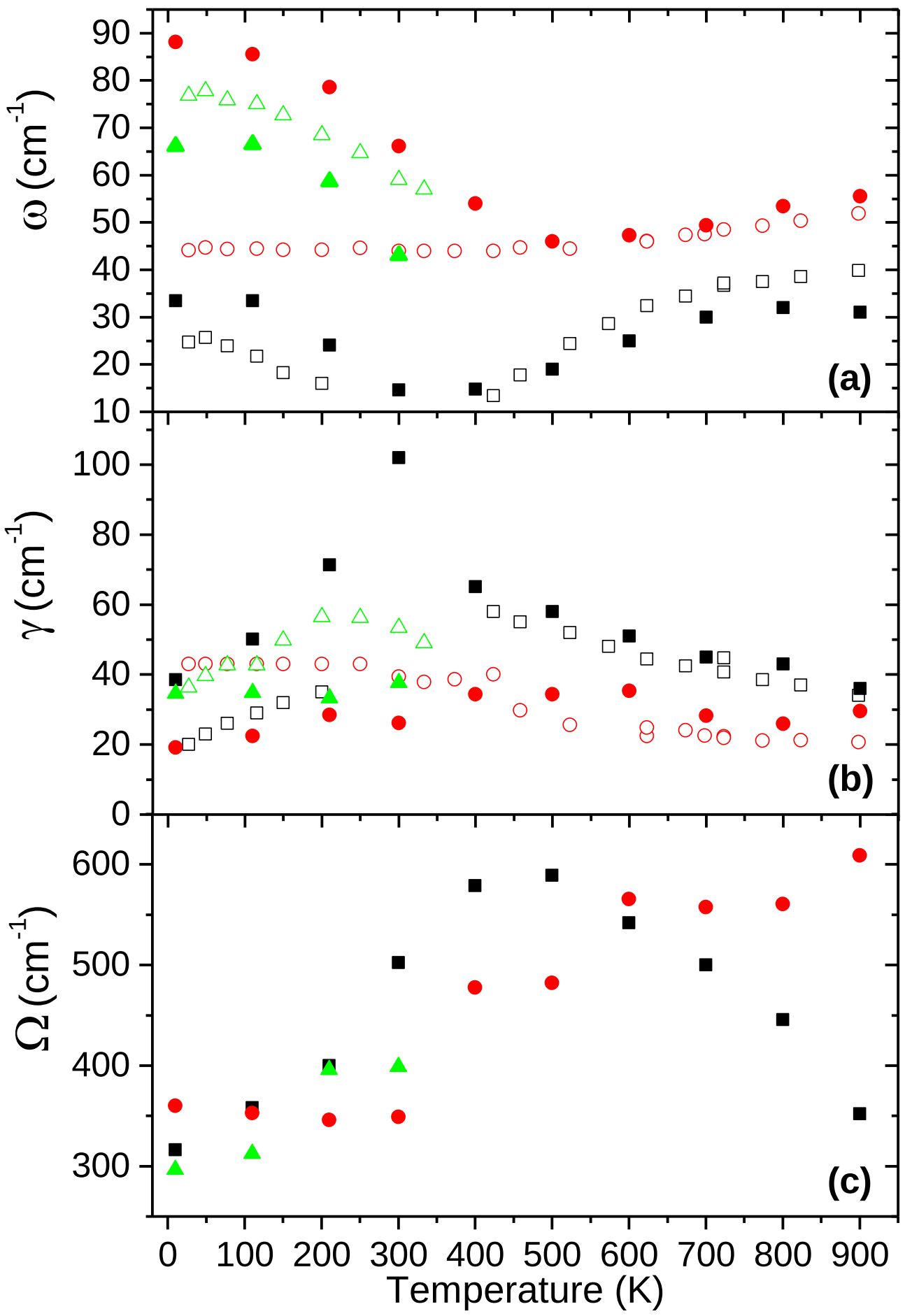

FIG. 6. Temperature dependences of the parameters of the three lowest-frequency TO phonon modes (marked by the arrows in the bottom of Fig. 3). Full symbols stand for the parameters from model (ii) in Figs. 4 and 5, open symbols stand for those from the fits to HRS data [17] – open symbols. The mode plasma frequency $\Omega$ is related to the oscillator strength $\Delta\varepsilon_{HO}\omega_{HO}^2$ (Eq. 3) by $\Omega = (\Delta\varepsilon_{HO}\omega_{HO}^2)^{1/2}$ [24].

The inspection of HRS spectra indicates that there might be an important coupling between the $F_{2g}$ mode and SM [17], which prevents their crossing at high temperatures. Therefore in Fig. 5 we have also included imaginary coupling [33] into our latter two-mode IR fits at 400 – 900 K. We see that the fit quality using the coupled oscillator formula (model (iii)) is improved and is comparable to the fits with model (i). Since it is well-known that the same spectra may be obtained with real as well as imaginary coupling constants [33], we also recalculated the same spectra using the real coupling constants by obtaining the same fitted spectra. However, the bare mode frequencies from both types of couplings do not correspond to the temperature dependences suggested from the HRS [17] and do not indicate SM behavior of any of the bare modes. In fact, the effective softening of the lower-frequency dressed mode in the coupled–mode approach appears only due to a strong increase of the coupling constant, which is not physically appealing and does not correspond to the HRS picture. So, there are appreciable differences among the temperature dependences of the parameters in the IR and HRS spectra above 300 K, which show that the agreement between the HRS and IR spectra is not quite perfect, see the discussion below. At 300 K and below, both HRS and our THz-IR spectra reveal three modes below ~100 $cm^{-1}$. As seen from Fig. 4, the fits with model (ii)) are also not quite perfect, particularly at 300 K. Even a trial coupling of the two lower-frequency modes (not shown) has not improved the fits sufficiently.

## V. MODELLING WITH EFFECTIVE MEDIUM APPROXIMATION

As a further attempt of evaluation, we used an alternative method, fitting procedure based on the EMA modelling used in Ref. [25] at room temperature, assuming that the slowly fluctuating or frozen PNDs are dielectrically anisotropic with a random orientation of the local polarization $\boldsymbol{P}_l$. Using this modelling, the higher-frequency (~45 $cm^{-1}$) mode at $T \geq 300$ K is assigned to the $A_1$-component of the SM doublet rather than to the $F_{2g}$ mode. For fitting the spectra at $T < 300$ K we had to account for the weak third mode, not seen at 300 K in [25], assigned in Ref. [17] to the $F_{2g}$ mode. We first added this mode only to the $A_1$-response ($\boldsymbol{E} \parallel \boldsymbol{P}_l$), which is usually more active in the Raman response. However, for good fits below 300 K we had to add an extra mode near 70 $cm^{-1}$ also to the $E$-response, which could be assigned to the second component of the weakly split $F_{2g}$ mode. For EMA modelling we have used two models with a threshold-limiting or weak percolation of the $A_1$-response [26,31]: Bruggeman model used in [25] (at the percolation threshold) and Lichtenecker model with a small positive $\alpha$ exponent (weak percolation) [31]. Recently, we have successfully used similar models for modelling the IR reflectivity spectra of alumina ceramics with highly anisotropic grains using the knowledge of the sapphire single-crystal spectra [34]. The fitted reflectivities with Bruggeman-EMA modelling (model (iv)) are shown in Fig. 7, compared with the THz data and model (i) fits used in Figs. 1-3. The fit quality by using the Lichtenecker model (not shown here) is quite comparable to that by Bruggeman, only the small reflectivity features near 430 $cm^{-1}$ due to so-called geometrical resonance [25] are better resolved in the Bruggeman model.

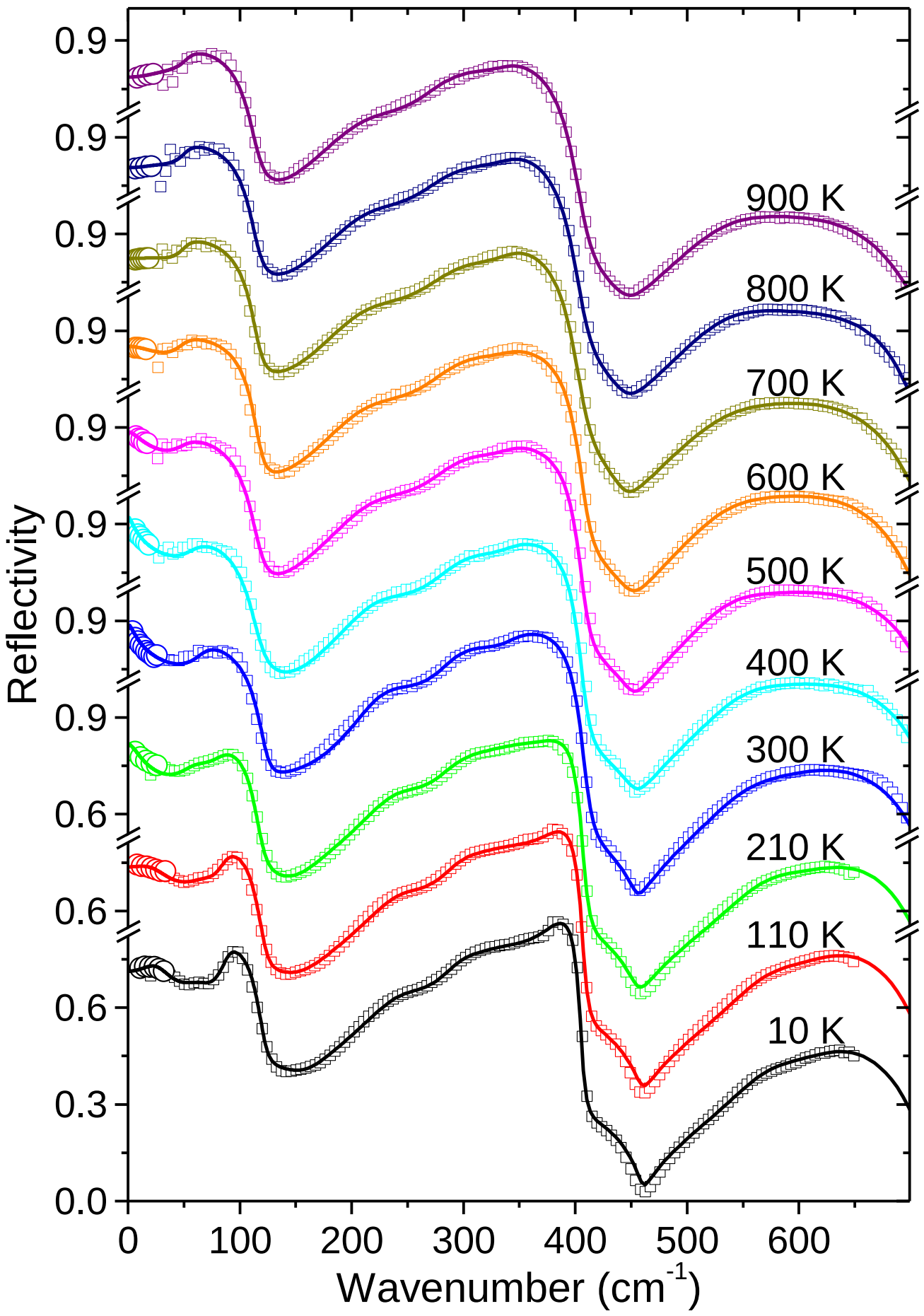

FIG. 7. IR reflectivity fits using the Bruggeman-EMA model (iv) in the whole temperature range – full lines, as compared with the THz data (circles) and IR reflectivity data (squares).

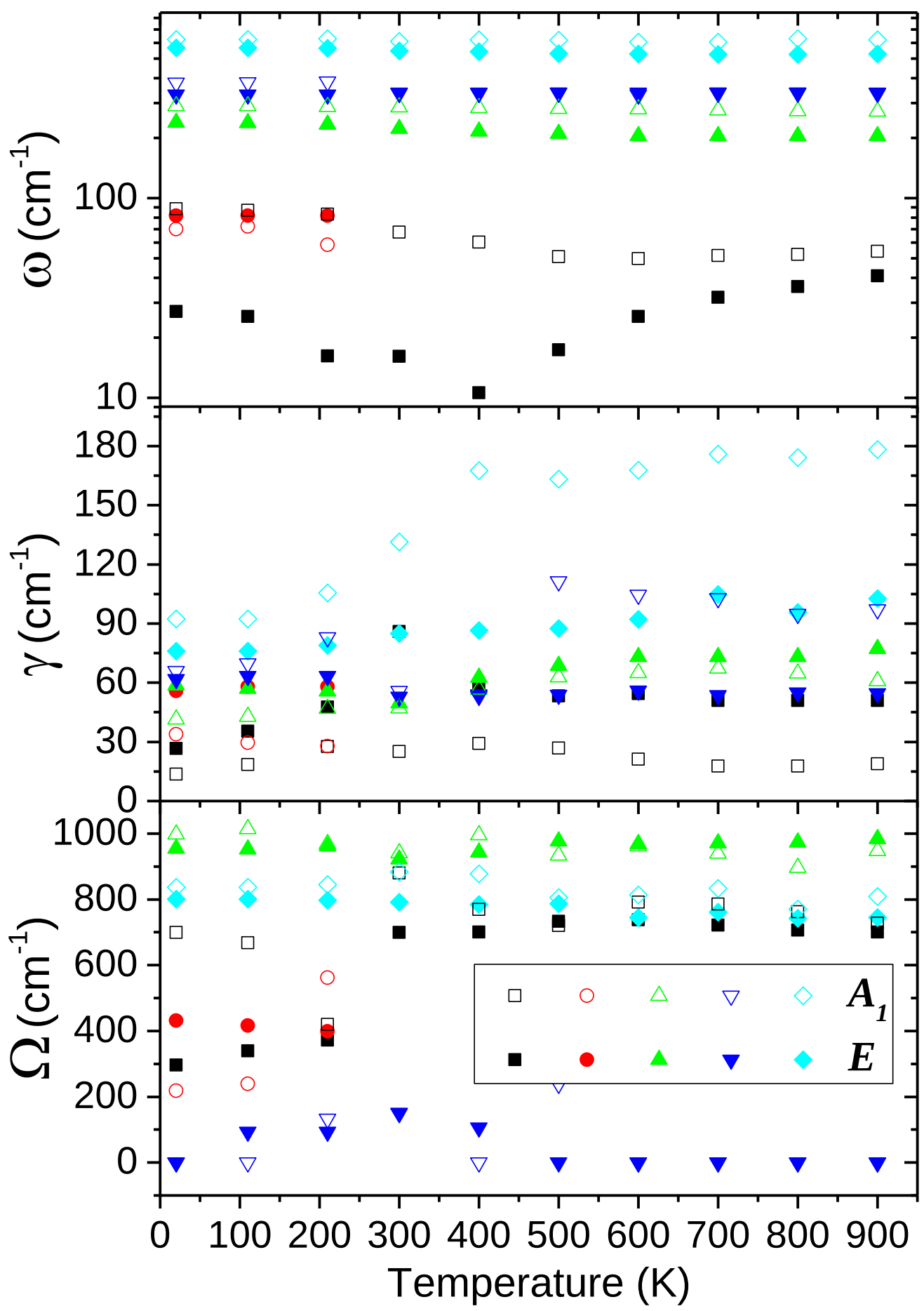

FIG. 8. Temperature dependences of the parameters from the Bruggeman-EMA fits (model (iv)) shown in Fig. 7. The mode frequencies are shown in the log-wavenumber scale for better distinguishing the SM behavior. Open symbols – $A_1$ spectra, full symbols – $E$ spectra. Modes below 100 cm$^{-1}$ represent the Last mode (SM), modes between 200 and 400 cm$^{-1}$ the Slater mode and modes above 500 cm$^{-1}$ the Axe mode.

Temperature dependences of the fitting parameters in both the $A_1$ and $E$ spectra are shown in Fig. 8. In Figs. 9(a) and 9(b) we plot the temperature dependent complex dielectric spectra of the $A_1$ ($\boldsymbol{E} \parallel \boldsymbol{P}_l$) and $E$ ($\boldsymbol{E} \perp \boldsymbol{P}_l$) symmetry, respectively.

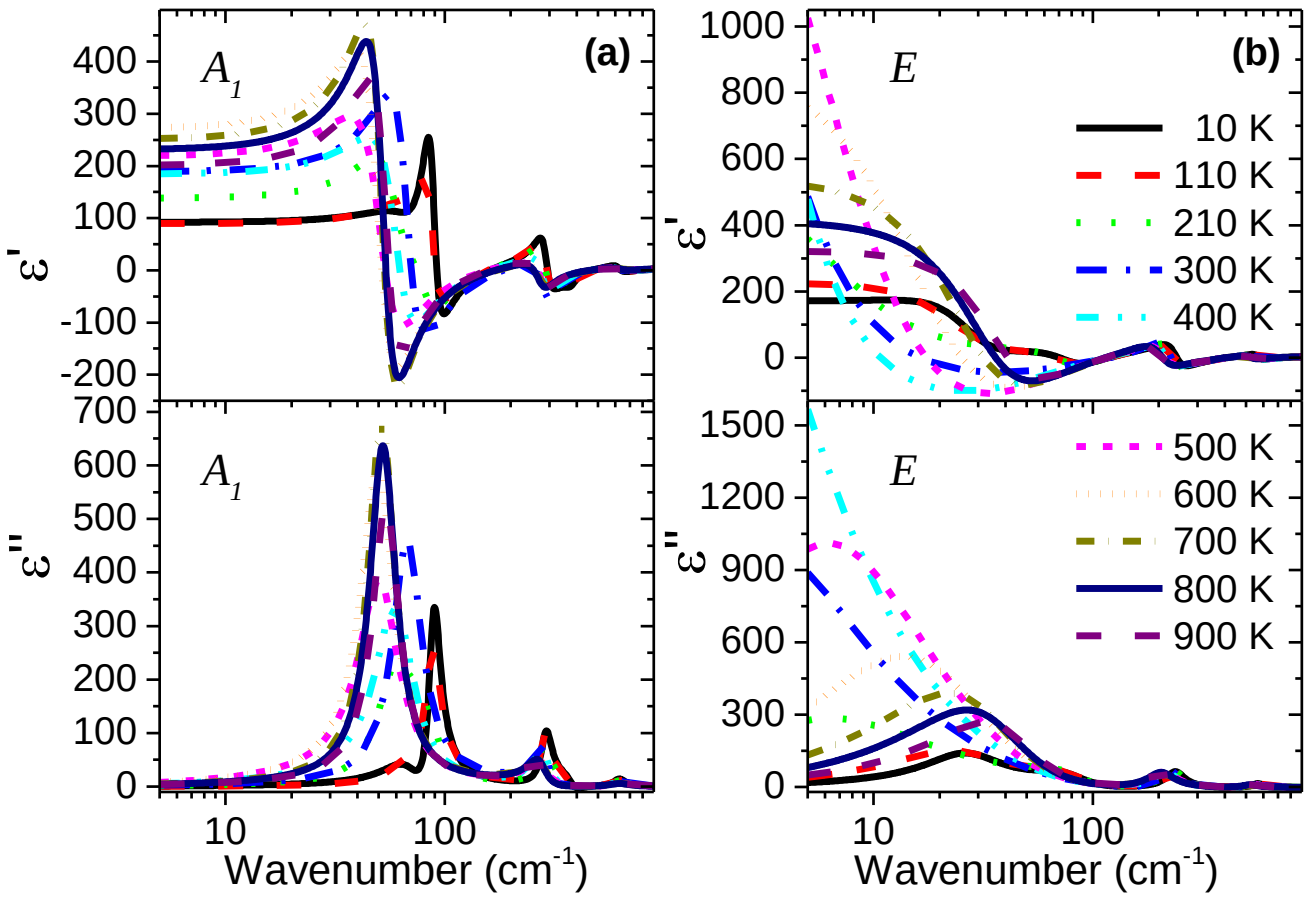

FIG. 9. Complex dielectric spectra of (a) $A_1$ ($\boldsymbol{E} \parallel \boldsymbol{P}_l$) and (b) $E$ ($\boldsymbol{E} \perp \boldsymbol{P}_l$) symmetry as evaluated from the fits in Figs. 7 and 8 (model (iv)) at selected temperatures (log-wavenumber scale).

In Fig. 10 we plot the temperature dependence of the static permittivity from our dielectric spectra of model (iv) (Fig. 9). It appears that the dielectric anisotropy is rather pronounced, the $E$-response ($\perp \boldsymbol{P}_l$) being higher and particularly the softening and the dielectric maximum at 400 K appear only in the $E$-response. To illustrate better the pronounced SM anisotropy between the $A_1$ and $E$ response, in Fig. 11 we plot the conductivity spectra of the SM for both components for all the temperatures. One can see that the $E$-symmetry overdamped soft mode has the lowest frequency near 400 K, whereas the $A_1$-symmetry SM exhibits only a rather shallow minimum near 600 K.

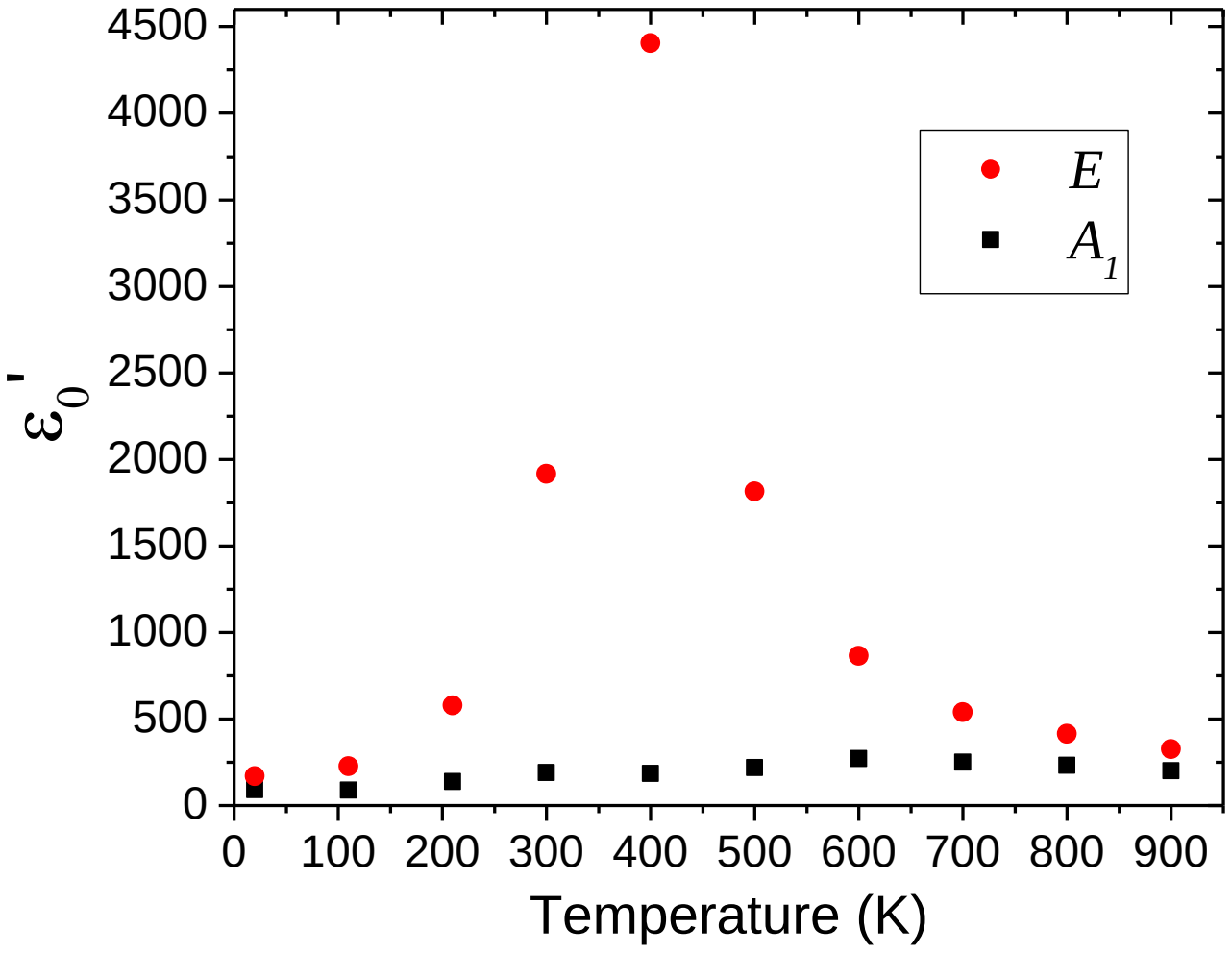

FIG. 10. Temperature dependence of the static permittivity calculated from the model (iv). Circles and squares correspond to the $E$ and $A_1$ response, respectively.

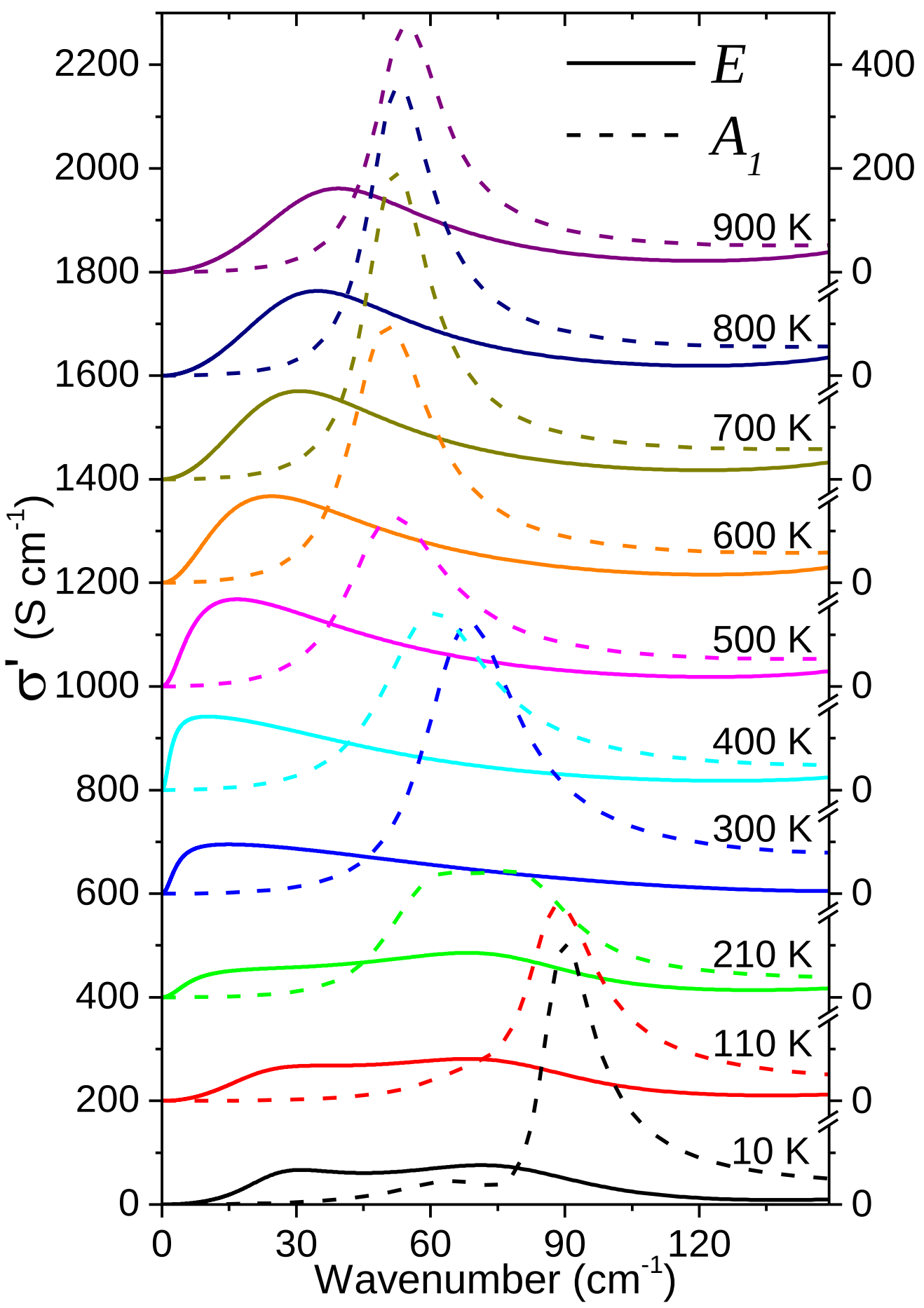

FIG. 11. Temperature dependence of the conductivity spectra for the SM components of the $A_1$ (dashed) and $E$ response (full lines) from model (iv). Each spectrum at the corresponding temperature is shifted by 200 S cm$^{-1}$ with respect to the previous temperature and the values on the left-hand y-axis match the spectra at 10 K, while on the right y-axis the proper scale is adapted for all the spectra.

## VI. MICROWAVE AND HIGH-FREQUENCY DATA

In Fig. 12 we have collected our earlier dielectric data [19,23] and the new THz data between $10^6$ and $10^{12}$ Hz (MHz – THz) at several temperatures between 100 and 500 K and their fits using Eq. 3 with one or two Cole-Cole relaxations. The fits include also the lower-frequency *E*-component of the SM, fitted by a single (mostly overdamped) harmonic oscillator, called CM in our previous papers [19,22]. The resulting temperature dependences of the fitting parameters are plotted in Fig. 13 with a small constant $\varepsilon_{ph}$ of ~15-30. From Fig. 12 it is seen that the fit quality, particularly at lower temperatures, is not very satisfactory. It appears that broader spectra with higher values of $\alpha$ for both Cole-Cole relaxations would give better fits, but their (unphysical) contribution to the THz range would spoil the fits in the CM region. Therefore here we prefer the realistic fits in the CM range of $10^{11}$-$10^{12}$ Hz with a negligible influence of the lower-frequency relaxations. In Fig. 13 we also plotted the rough fits of the two relaxation frequencies using the Vogel-Fulcher dependence for the lower-frequency relaxation and Arrhenius dependence for the higher-frequency one, in agreement with the literature in the lower-frequency relaxational range [2,20,23].

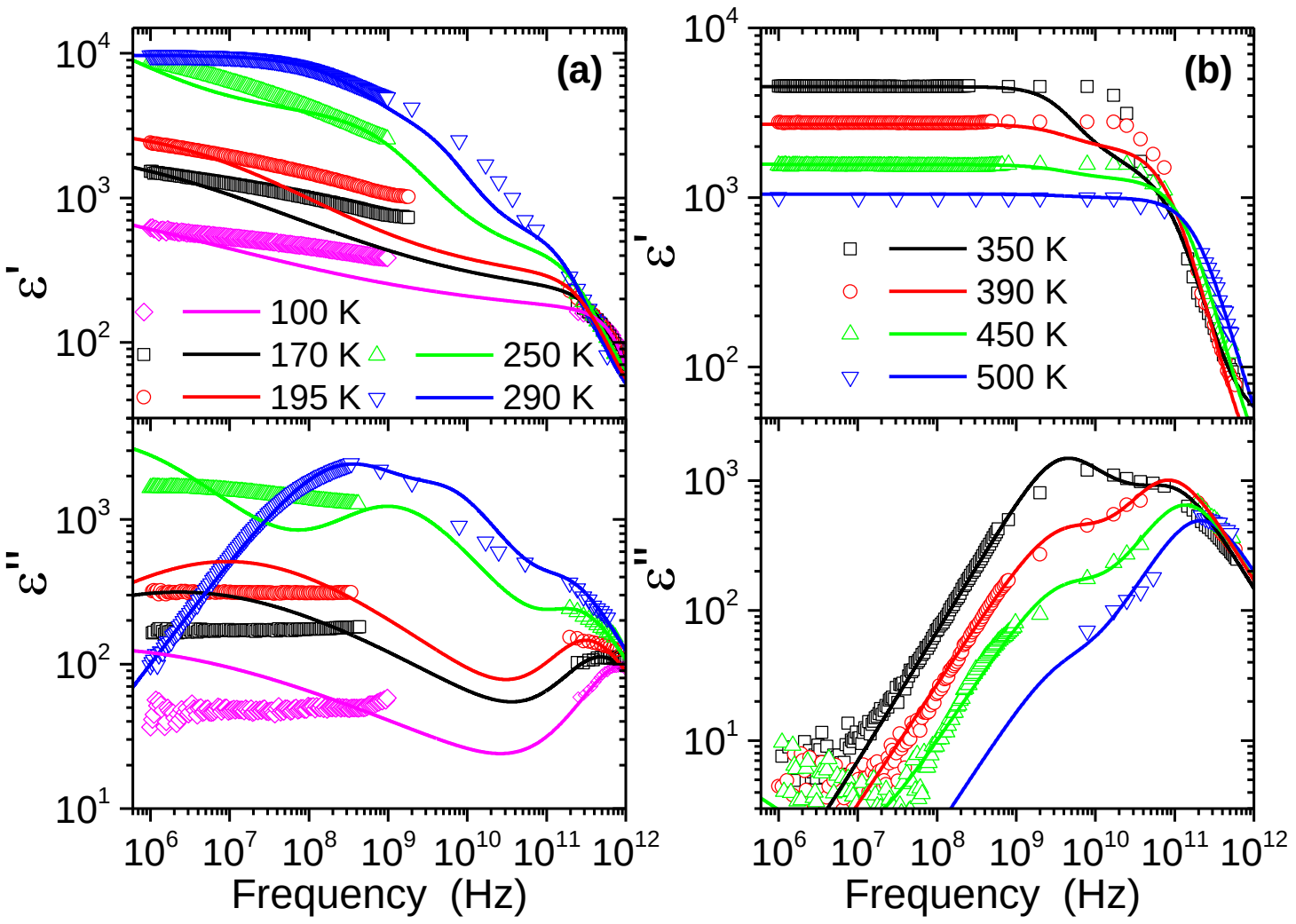

FIG. 12. Complex dielectric spectra of PMN above 1 MHz at selected temperatures in the 100–500 K range and their fits using Eq. 3: (a) temperatures below 300 K, (b) temperatures above 300 K.

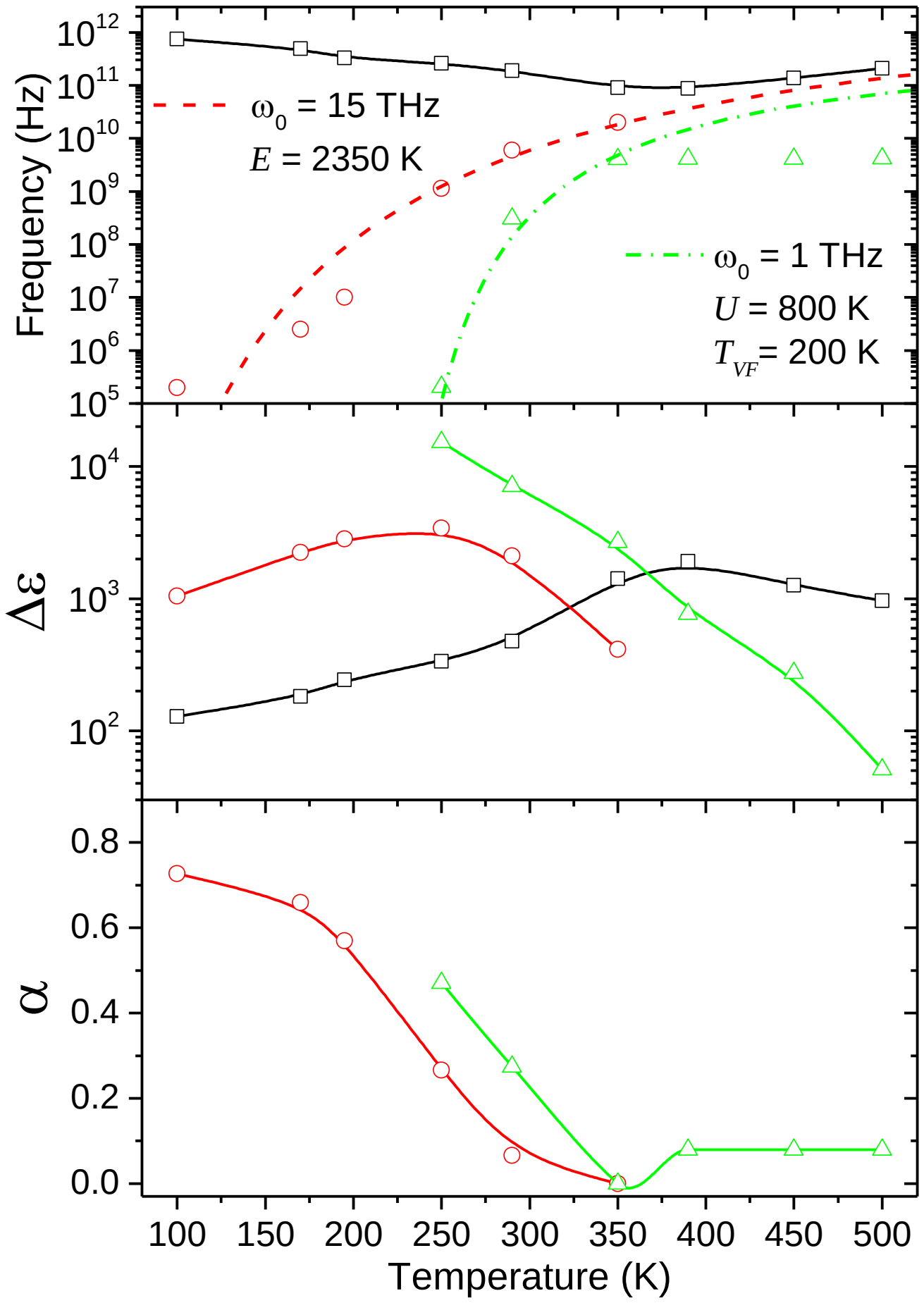

FIG. 13. Temperature dependences of the fitting parameters from Fig. 12 using Eq. 3. Frequencies of the Cole-Cole relaxations are for $\omega_{Ri}$, frequencies of the overdamped THz oscillator are given for $\omega_{HO}^2/\gamma_{HO}$. Dashed and dash-dotted lines are Arrhenius ($\omega = \omega_0\, exp(-E/T)$) and Vogel-Fulcher ($\omega = \omega_0\, exp[-U/(T-T_{VF})]$) fits of the temperature dependences of relaxation frequencies, respectively, with parameters shown in the figure. Only the temperature range ≤ 350 K, where both relaxations were resolved, was fitted. Full lines are guides for the eyes.

## VII. DISCUSSION

The good fits of the IR reflectivity in Fig. 1 using Eqs. 1 and 2 need 11 generalized oscillators, which is much more than expected for perovskites with a simple cubic structure, which have only 3 $F_{1u}$ IR active TO phonon modes. Nevertheless, the conductivity spectra in Fig. 3 show three main (slightly split) peaks, which can be clearly assigned to the 3 $F_{1u}$ modes, which, on increasing frequency, have predominantly the eigenvectors of so-called Last, Slater and Axe mode [24]. The weak extra mode at ~430 cm$^{-1}$ seen in Fig. 3 is known to be the geometrical resonance resulting from the EMA-based fits [25]. In Fig. 14 we plot the temperature dependences of the mode-plasma frequencies $\Omega_m$ of these 4 modes calculated from our fits (models (i) and (iv)) according to the formula [24]

$$\Omega_m^2 = \sum_k \Omega_k^2 \qquad \Omega_k^2 = \varepsilon_\infty \frac{\prod_j(\omega_{LOj}^2-\omega_{TOk}^2)}{\prod_{j\neq k}(\omega_{TOj}^2-\omega_{TOk}^2)} \qquad (6)$$

For our model (i), out of the 11 oscillators using Eq. 6, on increasing frequency the first 3 below 100 cm$^{-1}$ are involved into the Last mode, the next 5 modes between 150 and 400 cm$^{-1}$ into the Slater mode, the mode at around 430 cm$^{-1}$ is considered to be geometrical resonance (GR) [25] and the last two modes between 550 and 670 cm$^{-1}$ are involved in the Axe mode. As expected [24], the strongest mode is predominantly of the Slater type and the weakest one, which corresponds to the SM, is of the Last type. The slight temperature dependence of the Last mode-plasma frequency (particularly in Fig. 14 a, b) indicates some weak temperature dependence of the SM eigenvector. In Fig. 14 we show also the total mode-plasma frequency, calculated as

$$\Omega_{tot}^2 = \sum_{m=1}^{4} \Omega_m^2 \qquad (7)$$

which is expected to be temperature independent if the coupling of polar phonons with the electronic excitations can be neglected or is temperature independent. One can see that this is well satisfied for our spectra, which gives a good credibility to our data.

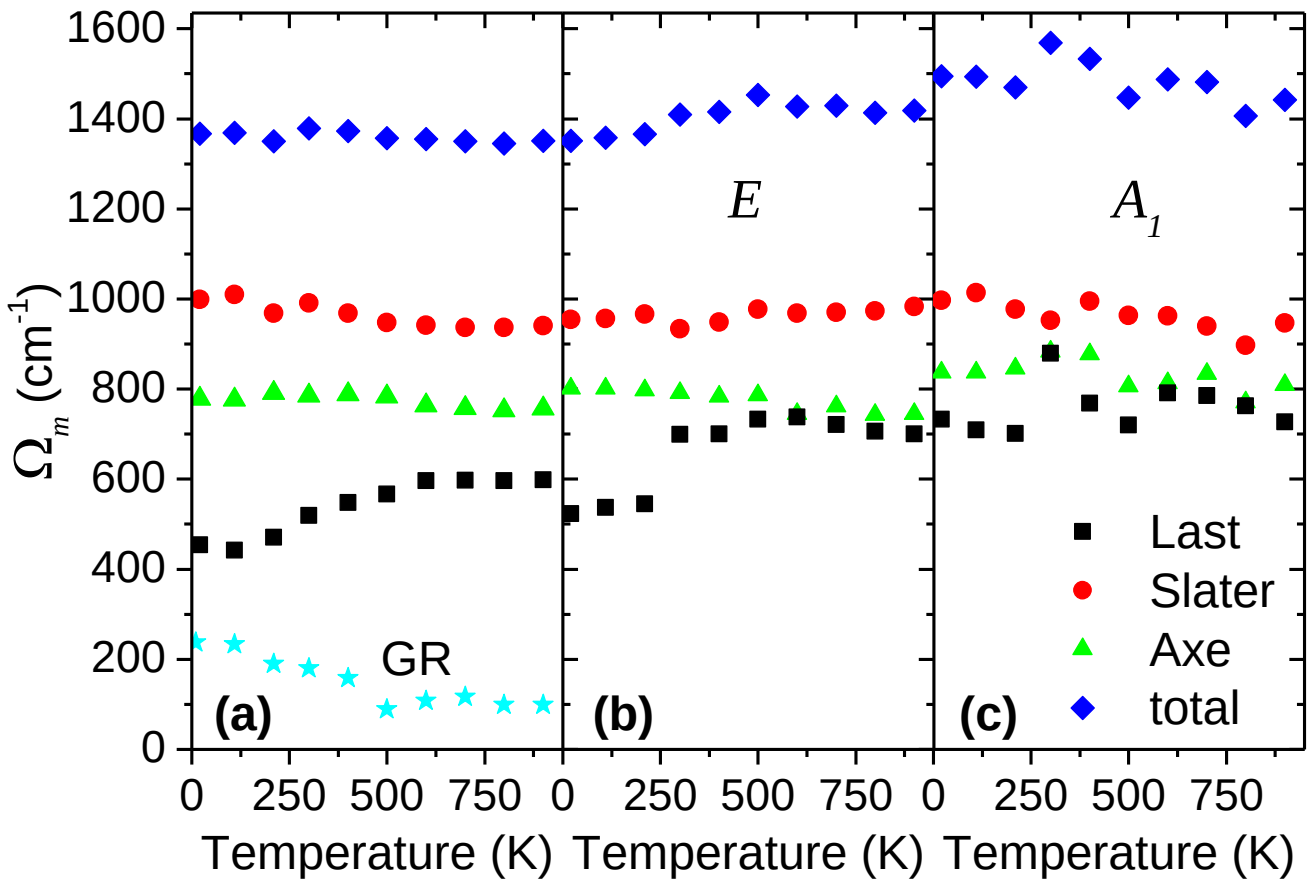

FIG. 14. Temperature dependence of the mode-plasma frequencies $\Omega_m$ of the main modes and the total plasma-mode frequency (from (a) model (i) and (b, c) model (iv)). GR stays for the geometrical resonance.

Comparison of IR fits with HRS in the SM frequency range below ~100 cm$^{-1}$ (Fig. 6) shows that the main differences appear at low temperatures. Even the EMA-based fitting (Fig. 8) does not reduce the difference between both types of experiments. The existence of the IR doublet in the 60-90 cm$^{-1}$ range was in fact indicated also by the raw HRS data, but due to their overlapping it was fitted with one oscillator only. More striking is the absence of any IR mode near 45 cm$^{-1}$, which is quite pronounced in HRS. One possible reason for the difference could be the different wavelength of the probing wave, which is in the order of 0.1-1 mm for the THz wave, but only 500 nm for the HRS wave, i.e. 3 orders of magnitude less. In the latter case of HRS, the size of sample nano-inhomogeneities (PNDs) might be not quite negligible compared to the wavelength. Moreover, HRS spectra may involve contributions of non-polar modes, therefore not active in IR spectra. Also, each spectroscopy probes different sample depth: a near surface layer in IR reflectivity (~ 1 μm), a thicker layer in the THz experiment (~40 μm), and in HRS the laser beam was focused relatively deep inside the bulk (~200 μm). Therefore any sample inhomogeneities in these scales below the sample surface, as observed in electric field by comparison of neutron and X-ray diffraction techniques [35], can also result in some differences between IR and HRS response. Finally, let us mention the difference between the studied samples: single crystal in the case of HRS [17] and ceramics [26] in the case of our THz experiment. However, our new low-temperature THz data agree well with our earlier data on a single crystal [21].

Let us now discuss the dielectric response below the THz range (Figs. 12 and 13). Even if the available high-frequency data are limited to the temperatures below 500 K, in agreement with the appearing deviations from the Curie-Weiss law [2,23,36], it can be accepted that the dispersion in this range emerges below the Burns temperature $T_B$ and can be connected with the appearance of dynamic PNDs, as suggested first time in Ref. [20]. On cooling from high temperatures, near $T^*$ the microwave dispersion appears to split into two parts which we fitted with two Cole-Cole relaxations which slow down and broaden on cooling. The dielectrically dominant is the lower-frequency relaxation which follows a fast slowing-down from the GHz range roughly according to the Vogel-Fulcher law with the freezing temperature $T_f \approx T_{VF} \approx 200$ K (see also Ref. [2] and references therein). This relaxation accounts for the frequency-dependent temperature maximum of the permittivity and dielectric losses. But in addition to it, the higher-frequency relaxation slows down and broadens from the 10 GHz range roughly according to the Arrhenius law, as also evaluated from the lower-frequency dispersion [2,37]. The higher-frequency relaxation persists above the MHz range with decreasing strength down to the lowest temperatures. It seems to be natural that below $T_f$ the losses can occur only due to fluctuations of the PND boundaries, so-called breathing of PNDs, since the $\boldsymbol{P}_l$ in the PND bulk is frozen. On the other hand, close but below $T_B$, where the local polarization $\boldsymbol{P}_l$ of the PNDs grows on cooling from very small values, the dominant dynamics is due to the flipping of $\boldsymbol{P}_l$, which must, however, slow down to zero near $T_f$. Therefore, to the first approximation, our two dispersion regions can be assigned to the breathing of PNDs and flipping of $\boldsymbol{P}_l$, respectively. This assignment, suggested already in Ref. [20], seems to be qualitatively justified independently of the microscopic picture of PNDs, particularly of their shape, size and whether their boundaries are sharp or diffuse. However, we stress that our EMA-modelling approach assumes that the dielectrically appreciably anisotropic PNDs occupy (with sharp boundaries) the whole sample volume with random orientation of $\boldsymbol{P}_l$, irrespectively of their size. On the other hand, concerning the relaxational dispersion of PNDs in relaxor perovskites, a mesoscopic theory based on the Landau approach close to the morphotropic phase boundary was suggested in Ref. [38], which explains well particularly the low-temperature response ($T < T_f$) of frequency-independent losses (within the limits of accessible frequency range) and their temperature dependence. The theory [38] is based on the assumption that the total dielectric response is due to a composite of frozen PNDs, which do not contribute to the relaxational response, and of dynamic dipoles at the PND boundaries, which can still flip between their possible orientations over some temperature independent statistical distribution of potential energy barriers.

High-frequency dynamics of PNDs in PMN was alternatively investigated using inelastic light scattering [14]. Below $T_B$, Koreeda et al. [14] observed an increase of the reduced intensity of the quasi-elastically scattered light, which followed power-law dependence in a broad frequency range of 1-2000 GHz. The relevant exponent $\mu$ (evaluated from a more restricted frequency range) decreased from 2 near $T_B$ to 1.33 at $T_f$ and remained constant below $T_f$. Koreeda et al. [14] used percolation theory for disordered systems and explained it by the percolation of polar nanoregions below $T_f$. If we assume that the underlying physical mechanism of inelastic light scattering (i.e dynamics of PNDs) is the same as for dielectric losses, the reduced intensity of the scattered quasielastic light multiplied by frequency (see Fig. 3 in Ref. [14]) should be proportional to the *AC* conductivity spectra $\sigma'(\omega) = 2\pi\varepsilon_0\omega\varepsilon''(\omega)$. In Fig. 15 we plotted our $\sigma'(\omega)$ spectra for selected temperatures and it is seen that our exponent $\mu$ evaluated from the range of ~$10^6$-$10^{11}$ Hz decreases from ~1.6 at 500 K to ~1 below $T_f$. The value $\mu$ = 1 (see all our data above the MHz range at 100 K) corresponds to frequency-independent loss $\varepsilon$''($\omega$) generally appearing in relaxors below $T_f$ (also known as the 1/*f* noise) [38], see Fig. 12(a). It can be obtained from a very broad (much broader than the experimentally probed frequency range) uniform distribution of Debye relaxation times, corresponding to Cole-Cole relaxation (Eq. 3) with $\alpha$ approaching 1 [39]. The maximum value $\mu$ = 2 corresponds to low-frequency tails of the Debye relaxation (as well as damped/overdamped harmonic oscillator) for the Cole-Cole limit $\alpha$ = 0. In our high-frequency spectra below 1 GHz it appears near ~350 K, see Fig. 15 as well as Fig. 13. In the microwave range near 350 K we obtained $\mu$ = 0.9. We see that our $\sigma$'($\omega$) spectra do not follow any universal power-law dependence and correspond rather to our phenomenological fitting models. Apparently, the assumption that the absorption mechanisms determining $\sigma$'($\omega$) are the same as those for the light scattering processes is not justified. Therefore we can conclude that our $\sigma$'($\omega$) spectra differ from the inelastic light scattering spectra of Ref. [14] and do not support any percolation picture.

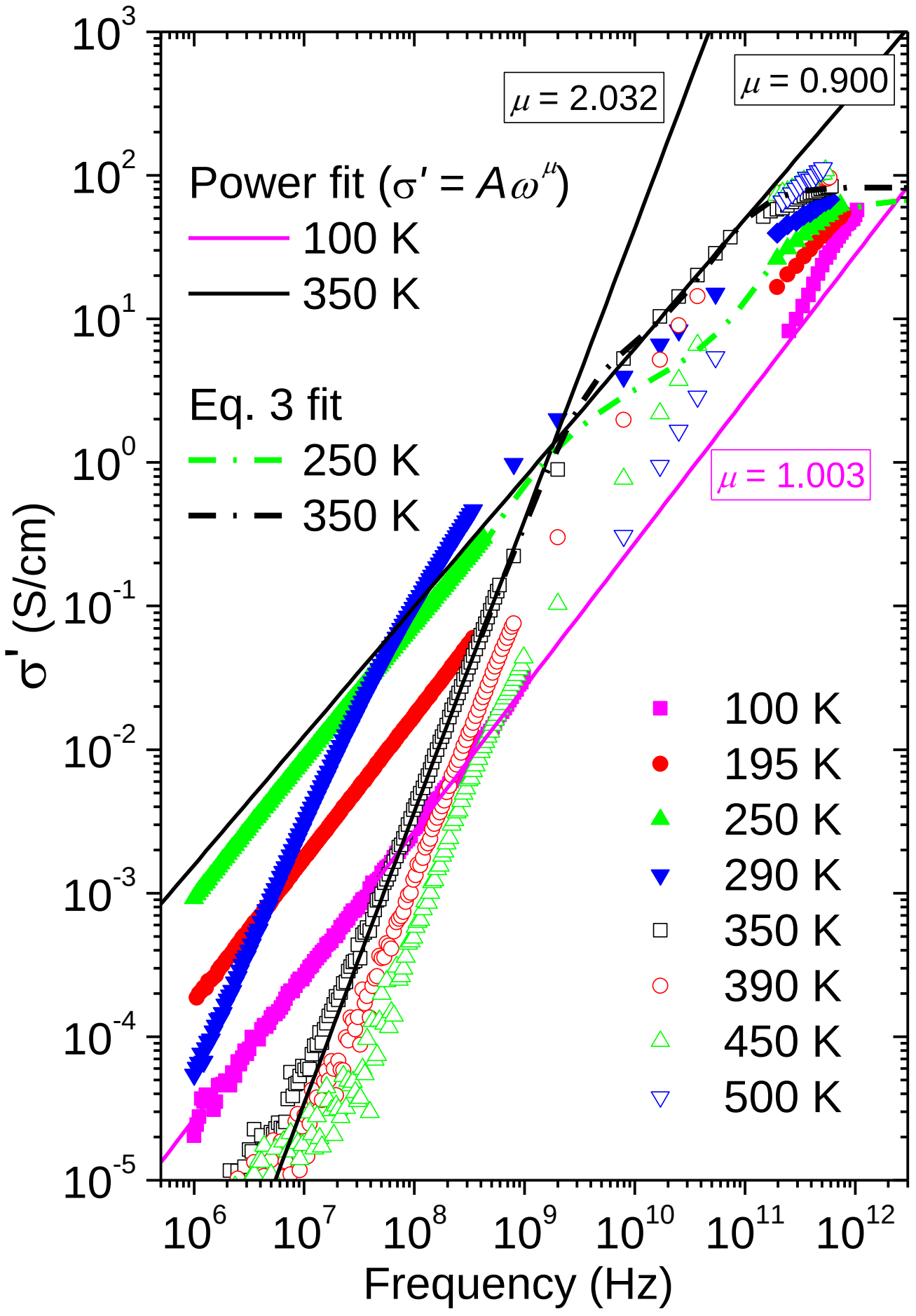

FIG. 15. Frequency dependences of the *AC* conductivity of PMN at selected temperatures calculated from the dielectric loss spectra in Fig. 12. Symbols correspond to the experimental data, solid lines for 100 and 350 K correspond to the power-law fits of the high-frequency and microwave data, dash-dotted lines correspond to our fits using Eq. 3 at 250 and 350 K.

Let us now discuss the SM behavior. From the dielectric as well as HRS experiment it appears that, on cooling from high temperatures, only the lower-frequency component of the SM ($E$-component in the EMA modelling) softens reaching the minimum effective loss-peak frequency of ~3 $cm^{-1}$ (~$10^{11}$ Hz) near $T^* \approx 400$ K contributing about 2000 to the effective static permittivity (see Fig. 13). Near this temperature the relaxational contribution (in the $10^{10}$ Hz range) has already comparable dielectric strengths and at lower temperatures it dominates in the dielectric spectra. This masks any anomaly or maximum in the low-frequency dielectric permittivity near $T^*$. The fact that only the $E$-symmetry SM component is softening, which accounts for the response perpendicular to $\boldsymbol{P}_l$, can be understood from softening of the multi-well potential for the anharmonic vibrations of the off-centered Pb ions, which essentially contribute to the SM eigenvector, mostly in this plane perpendicular to $\boldsymbol{P}_l$.

Let us more discuss the softening of the $E$-response. It is clearly reasonable that the maximum in the corresponding permittivity $\varepsilon'_{0\perp}$ (Fig. 10) is much higher than the maximum in the effective static response without relaxations (Fig. 13), since the $A_1$-response does not show any anomaly. In Fig. 16 we plot the temperature dependence of the inverse permittivity $1/\varepsilon'_{0\perp}$ and compare it with the corresponding temperature dependent frequency of the $E$-SM response. One can see a nearly perfect agreement with the Curie-Weiss (C-W) law $\varepsilon'_{0\perp} = C/(T-T_C)$ with the C-W constant $C = 1.7 \times 10^5$ K and Curie temperature $T_C = 380$ K.

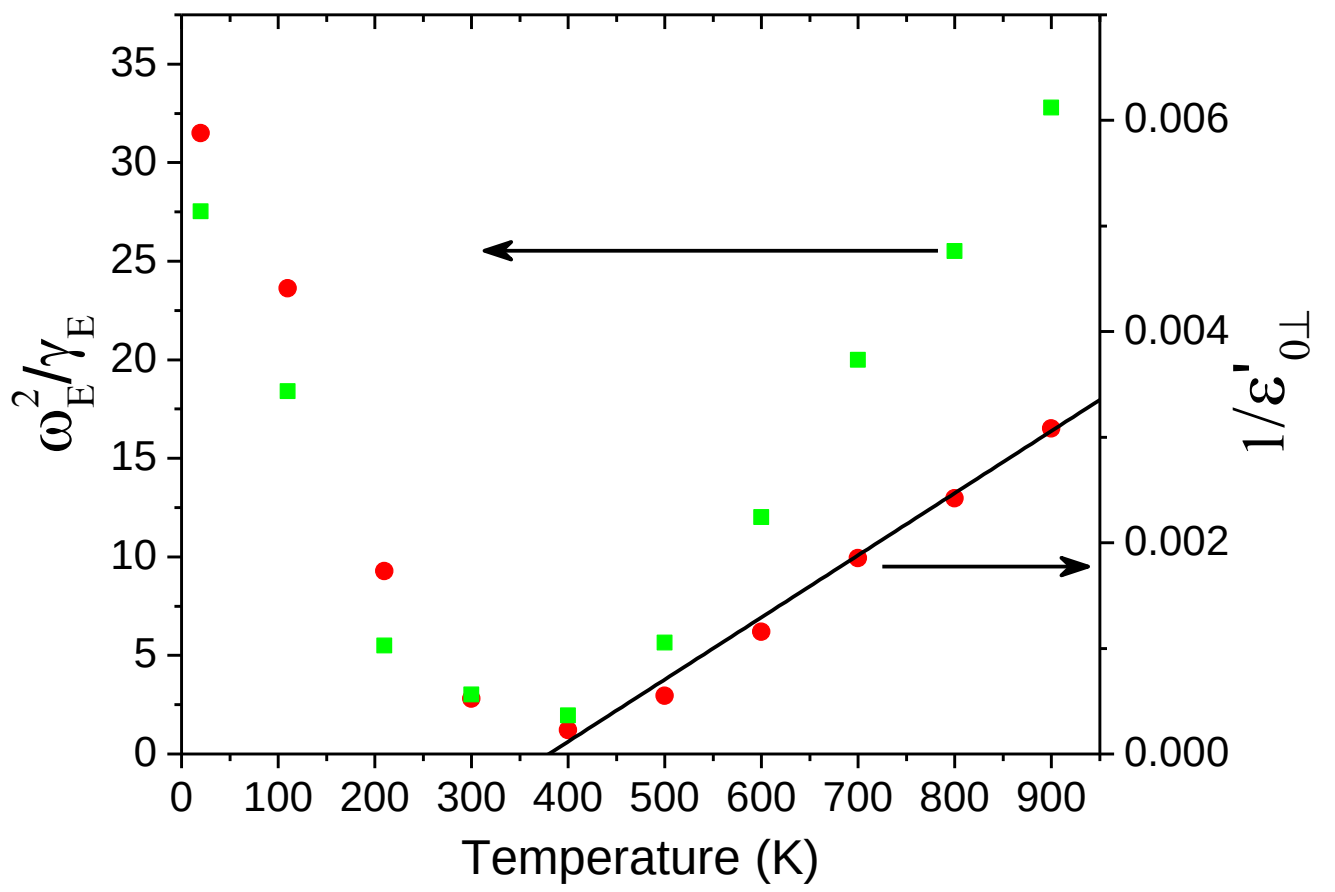

FIG. 16. Temperature dependence of the reciprocal static permittivity $1/\varepsilon'_{0\perp}$ (full circles) from Fig. 10 fitted with the Curie-Weiss law $\varepsilon'_{0\perp} = C/(T-T_C)$ ($C = 1.7 \times 10^5$ K, $T_C = 380$ K) (full line) and compared with the temperature dependence of the overdamped $E$-component of the SM characterized by $\omega_E^2/\gamma_E$ – full squares.

The local ferroelectric transition in PNDs, which concerns just the $\varepsilon_\perp$ response perpendicular to $\boldsymbol{P}_l$, appears to be of the second order within the limits of experimental accuracy and assuming justification of the Bruggeman EMA modelling. Also, the $E$-SM frequency $\omega_E^2/\gamma_E$ (corresponding to the peak in $\varepsilon''_\perp(\omega)$) is well proportional to $1/\varepsilon_\perp$, as expected for a classical SM-driven ferroelectric transition. The C-W parameters are in good agreement with $C = 1.25 \times 10^5$ K and $T_C = 398$ K obtained from the high-temperature (up to 800 K) low-frequency ($10^5$ Hz) permittivity fitting [36], where deviations from the C-W behavior appear below ~600 K. The latter slightly smaller $C$ value is in agreement with our EMA approach, which assumes a weaker divergence of the effective response due to the mixing of the critical $E$-response with the non-anomalous $A_1$-response along $\boldsymbol{P}_l$, and the deviations from the C-W behavior below ~600 K are due to the appearance of relaxational dispersion caused by the dynamical PNDs, also in agreement with our results.

Let us discuss our EMA-fitting results (model (iv) indicating that near $T^*$ the PMN possibly undergoes a local ferroelectric instability within the PNDs. For example, assuming that the PNDs have uniaxial polar symmetry with rhombohedral symmetry (as ferroelectric domains with space group *R3m*) and local polarization along (111) (*c*-axis of the rhombohedral phase) [40,41], above $T^*$, the local symmetry of PND could be still lower below $T^*$, either monoclinic (point group *m*) or just point group 1 (no symmetry element). It appears that the low-temperature structural data [40,41] do not exclude such a structure, but they indicate that the regions among the PNDs with inhomogeneous polarization could be quite broad, comparable to their diameter. This feature, of course, is not described by our model (iv).

In Fig. 17 we compare the temperature dependence of our SM frequencies with those obtained from different relevant experiments (HRS [17], Raman [3,13], INS [3,6,8,9]). We plot the oscillator frequency $\omega_0$ if the oscillator is underdamped with $\gamma < \omega_0$ and $\omega_0^2/\gamma$ if $\gamma > \omega_0$, since in the latter case this quantity can be experimentally more accurately determined (as the CM halfwidth from the scattering experiments). In the case of Raman data [3,13], only the lowest phonon mode can be evaluated, since at higher frequencies the hard $F_{2g}$ mode is dominating. One can see that each type of experiment yields somewhat different data with different softening. Our data are in the best agreement with those from HRS, showing the same minimum SM frequency near $T^* \approx 400$ K. Particularly, the Raman data [3,13] yield the temperature of minimum SM frequency much lower, near 270 K. As seen from Fig. 17, the available INS data on SM do not reveal the $T^*$ temperature either and give no information about the *E*-symmetry SM component below $T^*$.

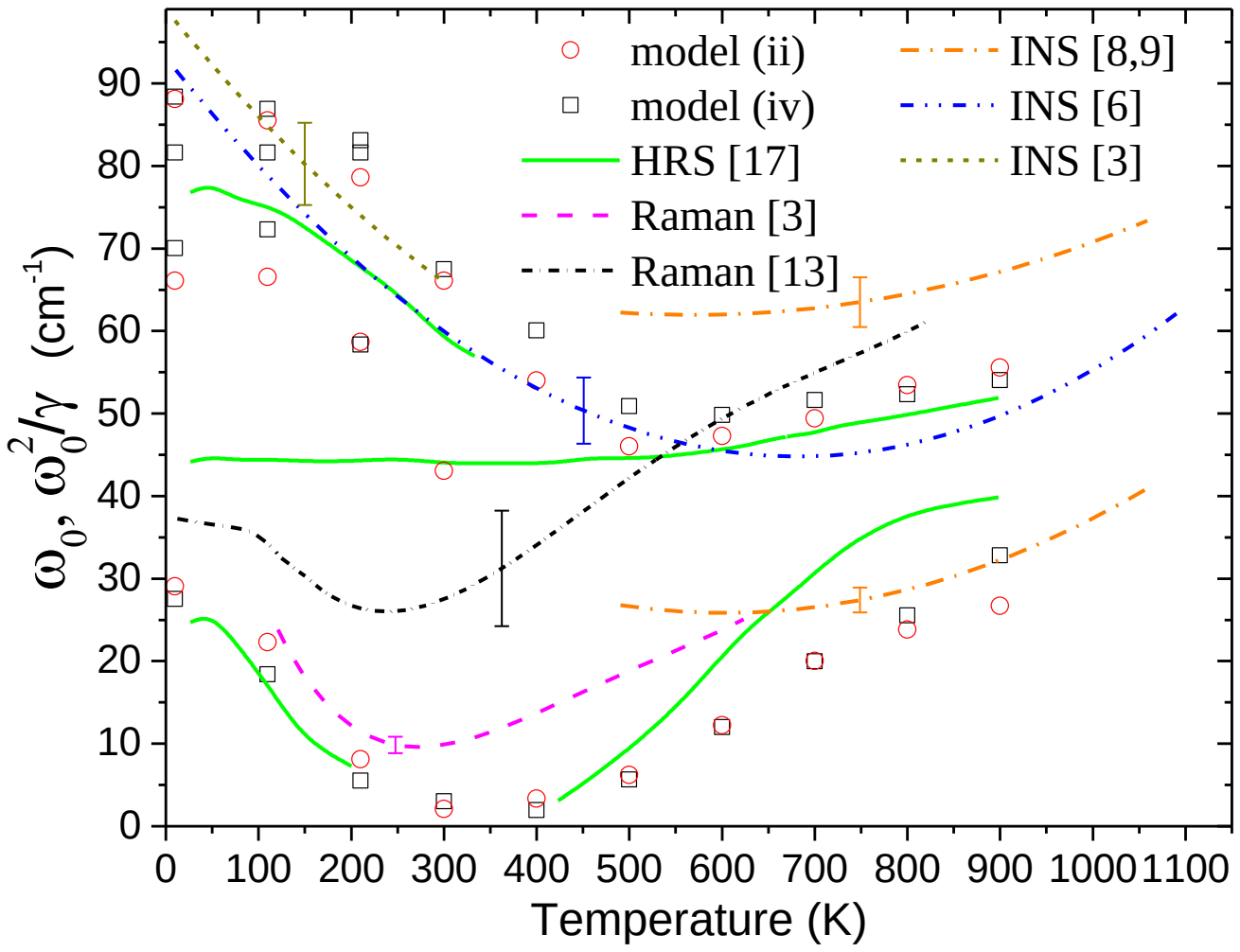

FIG. 17. SM frequencies from our THz-IR data (symbols) compared with other experiments: HRS [17], Raman [3,13], INS [3,6,8,9], where we have substituted the data points by some averaged lines with error bars indicating the data scattering. In cases of the lowest-frequency IR, HRS and Raman modes the oscillator frequencies $\omega_0$ and $\omega_0^2/\gamma$ are plotted for underdamped and overdamped modes, respectively. There the oscillators are mostly overdamped; underdamped modes are only at 10 K for the EMA-based fits, below 50 and above 800 K for the HRS data and below 200 and above 470 K for the Raman data [13]. All INS data are for the underdamped modes.

It is now better understandable why the softening of the SM seen in other available experiments is not complete not showing features expected for a standard ferroelectric transition. For the effective response it might be understood as due to the mixing of the soft $E$ with the hard $A_1$-response. Since the size of PNDs is substantially smaller than the wavelengths of IR and optical probes, also these experiments can probe only an effective response based on EMA, averaged in a specific way over all orientations of the anisotropic PNDs.

Let us now comment on the assignment of the higher component of the SM doublet seen above room temperature. The present IR reflectivity and THz spectra confirm the findings of the HRS experiments [17] in the sense that both modes persists well above the Burns temperature. Moreover, the temperature dependence of the corresponding mode frequencies and, in particular, their plasma frequencies from model (ii) in Fig. 6(c) suggest the scenario of the avoided mode crossing and consequent mode eigenvector mixing, similar to that inferred from HRS [17], even if our coupled-mode fit failed. From that point of view, the interpretation in terms of a single, primary IR active SM, accidentally interacting with a weakly IR active hard mode represent the most appealing interpretation for both the IR and HRS data.

On the other hand, the present IR reflectivity and THz spectra were also fitted by the EMA model (iv) as in Ref. [25], and these fits turned out to be remarkably good in the entire temperature range. This model was based on the assumption that the lower and upper frequencies of the split polar bands correspond to the polarization fluctuations along and perpendicular to the $\boldsymbol{P}_l$ within the individual PNDs [25]. The results of this work, showing the clear signatures of the band splitting and the successful adjustment of the IR spectra to the EMA model even at temperatures well above the Burns temperature $T_B$ imply that the local anisotropy cannot be explained only by PNDs appearing only below $T_B$. Indeed, most of the current theories (see Ref. [42] and references therein) assume that the PMN relaxor above $T_B$ should behave as a usual paraelectric material.

There are, however, at least two circumstances that may have similar influence on the vibrations of polar modes in the temperature range above $T_B$ as the PNDs would have. Firstly, the chemical disorder on the perovskite B-sites and off-centered positions of the Pb ions even above $T_B$ [41] that certainly act as sources of the random electric fields. Secondly, dynamics of the SM at high temperatures still remains largely dissipative and of rather low frequency, which might have a similar effect as PNDs to polar vibrations. Since the dynamics of SM is strongly temperature dependent, it seems that the former aspect could be a more natural explanation. In fact, it has been convincingly argued that the random fields are important ingredients of the relaxor behavior of PMN [42,43]. So, even though we do not have a clear understanding of the link between the random fields and the two-component SM response of PMN, which shows features of the avoided crossing with a weakly polar mode, we think that these observations constitute an important clue to disentangling the puzzle of the outstanding relaxor dynamics of the PMN crystal.

## VIII. CONCLUSIONS

Our evaluation of the THz-IR data of PMN in the broad temperature range has shown that, even if the expected three main modes (with Last, Slater and Axe eigenvectors according to the increasing frequency) dominate in the dielectric function, the detailed fitting requires up to 11 generalized oscillators by using the factorized form of the dielectric function. An equally good fit can be achieved using Bruggeman EMA modelling with somewhat less fitting parameters, assuming that the PMN mesostructure consists of randomly oriented uniaxially anisotropic regions with two local dielectric functions, whose axes are given by the local dipole moments of PNDs or, perhaps, by the local random electric fields at high temperatures above $T_B$. The dielectric function perpendicular to the unique axis reveals a SM, which softens from high temperatures towards $T^*$ and the corresponding permittivity component follows a classical Curie-Weiss law $\varepsilon_\perp = C/(T-T_C)$ with $C = 1.7 \times 10^5$ K and $T_C = 380$ K. On the other hand, the dielectric response parallel to the unique axis reveals no appreciable softening from high temperatures, only some hardening on cooling below room temperature. It indicates a local ferroelectric transition near $T^*$, which implies freezing of the local polarization perpendicular to the (111) direction and further lowering of the local symmetry of PNDs.

Detailed comparison with the most recent HRS data [17] shows only small quantitative differences in the SM parameters and their temperature behavior. Comparison with other published experiments characterizing the SM behavior in PMN (Raman, Brillouin, INS) indicates some differences among the data, which can be caused by the lengthscale of the mesoscopic disorder in the PMN structure and its relation to the wavelengths of the probes and by different selection rules for absorption and scattering processes. The THz-IR data at high temperatures allow the interpretation in terms of a single paraelectric SM having an accidental anticrossing with a weakly IR active or disorder induced hard mode near 45 cm$^{-1}$. Nevertheless, the success of the EMA modelling has re-open the possibility that the higher-frequency SM component above room temperature belongs predominantly to the $A_1$ component of the SM, at higher temperatures possibly related to the response along the direction of the local random electric field. We believe that these results will stimulate further clarification of the relation between the random electric fields, PNDs and the two-component SM response in this canonical relaxor material.

The relaxational response below the THz range due to the PND dynamics (flipping of $\boldsymbol{P}_l$) emerges below the Burns temperature $T_B \approx 620$ K and on cooling dominates over the SM response so that no low-frequency permittivity anomaly is seen near $T^*$. The dispersion splits into two parts in the frequency spectra and strongly broadens on cooling so that below the freezing temperature $T_f \approx 200$ K only rather strong frequency-independent losses appear up to the GHz range. They are assigned to fluctuations of the PND boundaries (PND breathing) and remind a very similar behavior of the relaxor PLZT [38].

## ACKNOWLEDGEMENTS

The authors thank G. Trefalt for providing us the dense PMN ceramics studied in Ref. [26]. This work was supported by the LD15014 Project of the Czech Ministry of Education (MŠMT) and the Czech Science Foundation No. 15-04121S.